\documentstyle[twoside,fleqn,espcrc2,psfig]{article}
\newcommand{\cA}{{\cal A}}

\newcommand{\cD}{{\cal D}}
\newcommand{\cE}{{\cal E}}
\newcommand{\cF}{{\cal F}}
\newcommand{\cL}{{\cal L}}
\newcommand{\cM}{{\cal M}}
\newcommand{\cN}{{\cal N}}

\newcommand{\cS}{{\cal S}}
\newcommand{\cT}{{\cal T}}
\newcommand{\cW}{{\cal W}}
\newcommand{\cZ}{{\cal Z}}
\newcommand{\rg}{\sqrt{g}}
\newcommand{\epsfig}{\psfig}
\newcommand{\lm}{\lambda}
\newcommand{\p}{\partial}
\newcommand{\hf}{\frac12}
\newcommand{\bea}{\begin{eqnarray}}
\newcommand{\eea}{\end{eqnarray}}
\newcommand{\be}{\begin{equation}}
\newcommand{\ee}{\end{equation}}
\newcommand{\bt}{\begin{tabular}}
\newcommand{\et}{\end{tabular}}
\newcommand{\ba}{\begin{array}}
\newcommand{\ea}{\end{array}}
\newcommand{\eps}{\epsilon}
\newcommand{\la}{\langle}
\newcommand{\ra}{\rangle}

\newcommand{\del}{\delta}

\newcommand{\La}{\Lambda}
\newcommand{\om}{\omega}

\newcommand{\tr}{{\rm Tr}}
\newcommand{\R}{{\rm Re}}
\newcommand{\I}{{\rm Im}}
\newcommand{\Del}{\Delta}

\newcommand{\sltwo}{SL(2,{\bf Z})}
\newcommand{\foot}{\footnote}
\newcommand{\Mstring}{M_{\rm str}}
\newcommand{\AmS}{{\protect\the\textfont2
  A\kern-.1667em\lower.5ex\hbox{M}\kern-.125emS}}

\title{HOLOMORPHIC COUPLINGS IN STRING THEORY}

\author{Jan Louis
        \address{Martin--Luther--Universit\"at Halle--Wittenberg,\\
        FB Physik, D-06099 Halle, Germany}
       \thanks{Lectures presented by J.~Louis at the Trieste Spring School 1997}
        and 
        Kristin F\"orger
        \address{Sektion Physik, Universit\"at M\"unchen, \\ 
    Theresienstr.~37, D-80333 M\"unchen, Germany}}

\begin{document}

\begin{abstract}
In these lectures we review the properties of 
holomorphic couplings in the effective 
action of four-dimensional $N=1$ and $N=2$
closed string vacua. We briefly outline their role
in establishing a duality among (classes of)
different string vacua.
\end{abstract}

\maketitle
\setcounter{section}{-1}
\section{Introduction}

Holomorphic couplings of the (four-dimensio\-nal)
low energy effective action
play an important role in string theory. 
This is largely due to the fact that supersymmetry 
protects them against quantum corrections.
In other words, they obey
a non-renormalization theorem and this property 
considerably simplifies their computation in string theory.
Furthermore, a certain subset of the holomorphic couplings can 
be calculated exactly
and not only in a weak coupling (perturbative) expansion.
As a consequence such couplings have been
used to support some of the conjectured dualities between 
seemingly different four-dimensional string vacua.

In these lectures we review some of the (older) perturbative 
computations and outline their relevance for string duality.
In particular, lecture~1  recalls some basic facts about 
perturbative string theories. 
Lecture~2 is devoted to $N=1$ vacua of the heterotic string while 
lecture~3 focusses on $N=2$ vacua of both the heterotic 
and type II string.
Finally, in lecture~4 we discuss the heterotic--type II duality.
At various points we make contact with  other
lectures presented at this school by
K.~Intriligator, 
S.~Kachru, W.~Lerche, K.S.~Narain, R.~Plesser and
J.~Schwarz.

\section{Perturbative String Theory}
\subsection{String Loop Expansion}

In string theory the fundamental objects are one-dimensional strings
which, as they move in time, sweep out a two-dimensional worldsheet
$\Sigma$.
This worldsheet is embedded in some higher dimensional target space
which is identified with a Minkowskian space-time.
Particles in this target space  appear as (massless) 
eigenmodes of the  string  and their scattering amplitudes are 
generalized by appropriate  scattering amplitudes of strings.
Strings can be open or closed, oriented or unoriented
but in these lectures  we solely focus 
on closed oriented strings. 
(For an introduction to string theory 
see for example
\cite{GSW,lusttheisen,Peskin,Polchinsky}.)

String scattering amplitudes are built from
the fundamental vertex depicted in figure~1
which represents the splitting of a string or 
the joining of two strings.
(Time is running horizontally.)

\begin{figure}[htp]
\input{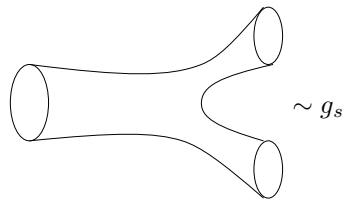}
\caption{Fundamental string interaction}
\end{figure}

The strength of this interaction
is governed by 
a  dimensionless string coupling constant $g_s$.
Out of the fundamental
vertex one composes all other possible 
string scattering
amplitudes, for example the four-point amplitude
shown  in figure~2.

\begin{figure}[htp]
\input{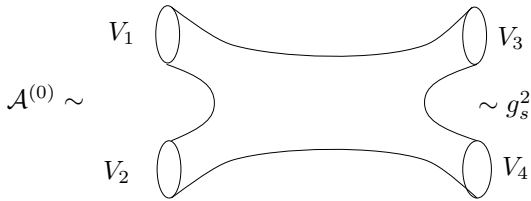}
\caption{Four-point  amplitude}
\end{figure}
 
The external `tubes' should be thought of as extending into
the far past and far future where the appropriate eigenstates
of the string are prepared. Technically this is achieved 
by the string vertex operators $V_i$. 

Obviously more  complicated scattering processes --
or equivalently more complicated worldsheets --  involving a non-trivial
topology can be built from the fundamental vertex.
The Euclideanized  version of any such worldsheet
is a two-dimensional Riemann surface
of a given genus $n$ where  
$n$ counts the number of holes in the worldsheet.
The total $N$-point string scattering amplitude $\cA$
is obtained by summing over all possible Riemann surfaces 
\be\label{sumA}
{\cal A}(V_1,\ldots, V_N, g_s) 
= \sum_{n=0}^{\infty}{\cal A}^{(n)}(V_1,\ldots, V_N, g_s)  ,
\ee
where $\cA^{(n)}$ denotes
the string scattering amplitude corresponding
to a worldsheet of genus $n$.
For example, a four-point amplitude of genus $n$ 
together with its $g_s$ dependence is displayed in figure~3.

\begin{figure}[htp]
\input{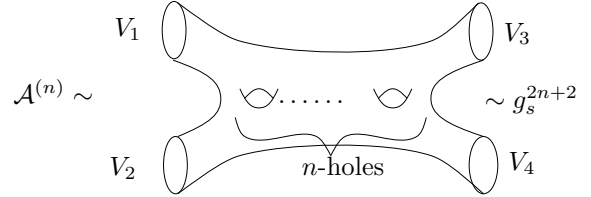}
\caption{Riemann surfaces of genus $n$}
\end{figure}

For an arbitrary 
$N$-point amplitude the $g_s$
dependence of ${\cal A}^{(n)}(V_1,\ldots, V_N, g_s)$
is easily found to be proportional to 
$g_s^{2n+N-2}$ but 
one commonly absorbs 
one power of $g_s$ into each vertex operator and defines 
$V_i' \equiv g_s V_i$.
Using the  rescaled vertex operators $V_i'$ one can eliminate the
$N$ dependence and define
\be
\cA^{(n)}(V_1',\ldots, V_N', g_s)\!
=\!  g_s^{2n-2} \cA^{\prime (n)}(V_1',\ldots, V_N')
\ee
where $\cA^{\prime (n)}(V_1',\ldots, V_N')$ 
no longer depends  on $g_s$.
As a consequence  eq.~(\ref{sumA}) turns into 
\bea\label{sumAs}
\lefteqn{{\cal A}(V_1, \ldots, V_N,g_s) =}\nonumber\\
& & \qquad \qquad \sum_{n=0}^{\infty}\  g_s^{2n-2}\ 
\cA^{\prime(n)}(V_1',\ldots, V_N')\ .
\eea
In this formula $g_s$ appears with a
power that coincides with  the (negative of the) 
Euler number $\chi$ of the Riemann surface 
\be\label{Euler}
\chi=\frac{1}{4\pi}\int_\Sigma \sqrt{h} R^{(2)}=2-2n\ ,
\ee
where $h$ is the world-sheet metric and $R^{(2)}$ the two-dimensional 
curvature scalar.

From eq.~(\ref{sumAs}) we  learn that 
expanding ${\cal A}$
in powers of the string coupling $g_s$ is equivalent to an
expansion in worldsheet topologies. 
This expansion can also be
interpreted as an expansion in the number of 
string loops and hence eq.~(\ref{sumAs})
is also known as  the string loop expansion.
For  $g_s<1$ the scattering amplitude $\cA^{(0)}$
 which corresponds to a 
worldsheet of genus $0$ or equivalently a sphere, 
is the dominant
contribution while higher genus amplitudes are 
suppressed by higher powers of  $g_s$. 
Our current understanding of string theory
does not fix the strength of the string coupling 
and leaves $g_s$ as a free parameter.
The regime  $g_s<1$ then 
defines what is referred to as 
`perturbative string theory'.

On the other hand, the strong coupling 
regime $g_s\ge 1$ was until recently
inaccessible
in that there were no non-perturbative methods available
for evaluating ${\cal A}$.
However, during the past two  years it was realized
that the strong coupling region of a given 
string theory often
can  be mapped to another weakly coupled,
`dual' string theory
and that most likely a non-perturbative formulation of string
theory not only contains strings but also 
other extended objects of higher dimension.
We briefly return to string dualities in lecture~4 but 
the recent developments have been nicely reviewed in the lectures by
J.~Schwarz and S.~Kachru. 

Despite the recent advances perturbative string theory has not gone
out of fashion yet. As we will see in the course of these lectures
the perturbative properties of the low energy effective actions
and in particular their holomorphic couplings  have played a vital part
in supporting the validity of some of the conjectured dualities.
Furthermore, the ultimate goal to connect string theory to
physical phenomena at the weak scale requires a much more detailed
knowledge about the perturbative sector as is currently available.
Therefore, in this first lecture we briefly 
summarize some basic facts of string perturbation theory.

\subsection{Conformal Field Theory}
So far we merely isolated the $g_s$ dependence of a string
scattering amplitude but did not compute the heart of the matter
$\cA^{\prime (n)}(V_1',\ldots, V_N')$ in eq.~(\ref{sumAs}).
A detailed review of the rules and techniques for computing
$\cA^{\prime (n)}(V_1',\ldots, V_N')$ is beyond the scope of these
lectures and we refer the reader to the literature for more details
\cite{GSW,lusttheisen,Peskin,Polchinsky,fms,GW,grosssl}.
For our purpose  we recall that 
the interactions of the string are governed by a two-dimensional 
field theory on the world-sheet $\Sigma$.
$\cA$ can be interpreted as an unitary scattering amplitude 
in the target space
whenever the two-dimensional field theory is
conformally invariant.
The conformal group in two dimensions is generated by the 
infinite dimensional Virasoro algebra whose generators $L_k$ obey
\be\label{virasoro}
[L_k,L_j]=(k-j)\, L_{k+j}+\frac{c}{12}(k^3-k)\, \del_{k+j,0}\ .
\ee
The constant $c$ is the central charge of the algebra which 
is constrained to vanish, i.e.~$c=0$.
Since we are discussing  oriented closed string theories the 
conformal field theory (CFT) 
is invariant under two separate conformal groups acting on the two
light cone coordinates $\sigma\pm\tau$.
($\tau$ is the two-dimensional time and $\sigma$ the space coordinate.)
In fact the entire partition function splits into two sectors
each of which carries a  representation of the Virasoro algebra.
These two sectors are commonly referred to as the left and 
right moving sector.

The different closed string theories
are defined by the amount of local worldsheet supersymmetry. 
The bosonic string has no worldsheet supersymmetry, while 
the superstring has one  supersymmetry in each  the left and right moving
sector; this is  called $(1,1)$ supersymmetry.
The heterotic string is a hybrid of the bosonic string and the
superstring in that it has one  supersymmetry only in the right moving
sector or equivalently $(0,1)$ supersymmetry.

The central charge in eq.~(\ref{virasoro}) has been normalized
such that a free (two-dimensional) boson contributes $c=\bar c=1$
and a (right moving) Majorana fermion  
has $c=\frac12$
($c\, (\bar c)$ denotes the central charges
of the right (left) moving sector). 
In addition to these `matter fields' of the CFT 
also 
the reparametrization ghosts of the worldsheet
contribute 
central charges $c_{\rm g}, \bar c_{\rm g}$.
If there is no supersymmetry on the worldsheet one finds 
$c_{\rm g}= \bar c_{\rm g} = -26$ while a locally supersymmetric
worldsheet has $c_{\rm g}= -15$.
For the total central charge to vanish these ghost contributions
have to be balanced by the central charge of the matter fields 
$c_{\rm m}, \bar c_{\rm m} $.
For the closed string theories this situation is summarized in
table~1. 

\begin{table}
\begin{center}
\begin{tabular}{|l|l|l|}\hline
string theories & supersymmetry & $(\bar{c}_{\rm m},c_{\rm m})$ \\\hline
bosonic string   & $\quad (0,0)$ & $(26,26)$\\
superstring      & $\quad (1,1)$ & $(15,15)$\\
heterotic string & $\quad (0,1)$ & $(26,15)$\\\hline
\end{tabular}
\end{center}
\caption{Worldsheet supersymmetry and central charges.}
\end{table}

The conformal symmetry ensures the consistency of the tree level
scattering amplitude $\cA^{(0)}$ 
but at higher loops  an additional
requirement has to be fulfilled.
The two-dimensional field theory also has to be invariant
under global reparametrizations of the higher genus 
Riemann surfaces.
At genus 1 (torus) the group of global reparametrizations
is the modular group $SL(2,\bf{Z})$ 
(some of its basic features are summarized in ref.~\cite{gpr}
and  appendix~A). 
$SL(2,\bf{Z})$ invariance severely constrains the 
partition function of a CFT and thus  the 
spectrum of physical states in the target space.
In particular it automatically ensures an 
anomaly free effective field theory in the target 
space \cite{ASNW}.

The bosonic string is afflicted with the problem of 
containing a (tachyonic) state with negative mass in its spectrum and 
the difficulty of  constructing fermions in space-time.
Therefore, we omit the bosonic string from our subsequent discussions
and only focus on the superstring and the heterotic string.
In both cases worldsheet supersymmetry requires the
presence of two-dimensional fermions in the CFT. 
Such fermions can have
different types of boundary conditions on the worldsheet: 
periodic (Ramond)  or  anti-periodic 
(Neveu-Schwarz). 
Modular invariance 
 requires to sum over all possible
boundary conditions of the worldsheet fermions and 
the states in the target space therefore arise in sectors
with  different fermion boundary conditions.
For example, in the heterotic string 
the $NS$ sector gives rise to space-time bosons 
while space-time fermions originate from
the $R$ sector.  
For the superstring there are worldsheet fermions
in both the left and right moving sector so that 
there are altogether four distinct sectors
$NS-NS$, $NS-R$, $R-NS$, $R-R$.
Space-time bosons now arise from the $NS-NS$ or  
$R-R$  
sector while 
space-time fermions appear in the $NS-R$ and $R-NS$ sector.

In order for a string to propagate in a 
$d$-dimensional 
target space (which should be identified with 
a Minkowskian  space-time) a subset of
the matter fields  of the CFT have to be 
$d$ free two-dimensional bosons 
together with the appropriate superpartners.
These free fields build up what is called the 
`space-time' (or universal) sector 
while the `left over' fields  can be an arbitrary 
(but modular invariant) interacting CFT 
called the `internal' sector. 
The central charges of the two sectors
are additive $c_{\rm m}=c_{\rm st}+c_{\rm int},$
where $c_{\rm st} (c_{\rm int})$ 
is the central charge of the space-time 
(internal) sector.
The balance of the central charges for a string propagating
in a $d$-dimensional space-time is  summarized in
table 2. 
\begin{table}\label{centchargd}
\begin{center}
\begin{tabular}{|l|c|c|}\hline
{} & $(\bar{c}_{\rm st},c_{\rm st})$ & $(\bar{c}_{\rm int},c_{\rm int})$ \\ \hline
\rule{0cm}{3ex}superstring & $(\frac{3}{2}\; d,\frac{3}{2}\; d)$ & $(15-\frac{3}{2}\; d,15-\frac{3}{2} \;d)$\\
heterotic       & $(d,\frac{3}{2}\; d)$ & $(26-d,15-\frac{3}{2}\; d)$ \\\hline
\end{tabular}
\end{center}
\caption{Balance of central charges in $d$ space-time dimensions.}
\end{table}

The space-time sector containing free two-dimensional
fields is more or less unique.
However, the interacting internal CFT is only
constrained by modular invariance and as we will see later 
also by the amount of space-time supersymmetry.
However, in most cases one finds 
a whole plethora of CFT which 
satisfy all constrains. Each of these CFT
together with their space-time sector is often referred to as
a string vacuum.

The dimension $d$ of space-time is completely arbitrary at this point.
The simplest case is to choose as many free fields as
possible which  corresponds to $d=10$ for both string theories.
In this case the
constraint from modular invariance is particularly strong and only leaves
four consistent closed string theories: 
the non-chiral type IIA, the chiral type IIB and the heterotic string with
a gauge group $E_8\times E_8$ or $SO(32)$.
Their  massless bosonic spectrum is summarized in  table 3.

\begin{table}
\begin{center}
\begin{tabular}{|l|l|l|}\hline
{}&$NS-NS$&$R-R$ \\\hline
IIA&$g_{MN},b_{MN},D$&$V_M,V_{MNP}$\\
IIB&$g_{MN},b_{MN},D$&$V_{MNPQ}^*,b_{MN}',D'$\\
Het&$g_{MN},b_{MN},D, V_M^{(a)}$&---\\\hline
\end{tabular}
\end{center}
\caption{Massless bosonic spectrum in $d=10$.}
\end{table}

In all four cases the $NS-NS$ sector 
contains the graviton $g_{MN}$, an  antisymmetric
tensor  $b_{MN}$ and a scalar $D$ 
called the dilaton.  
The heterotic string also has gauge bosons $V_M^{(a)}$
in the adjoint representation of either 
$E_8\times E_8$ or $SO(32)$.
These are the two anomaly free gauge groups in 
ten dimensions and this choice is also 
dictated by  modular invariance. 
The R-R sector of the type IIA string features an
Abelian vector $V_{M}$ and an antisymmetric 3-form
$V_{MNP}$. For  type IIB one finds an additional scalar $D'$, a
second antisymmetric tensor $b'_{MN}$  and   
a self-dual antisymmetric 4-form $V_{MNPQ}^*$.
The fermionic degrees of freedom are such that they
complete the ten-dimensional supermultiplets.
In type IIA one finds two spin-$\frac32$ gravitini of opposite
chirality (non-chiral $N=2$), in type IIB
there are  two gravitini of the same
chirality (chiral $N=2$),
and the heterotic string has one gravitino ($N=1$)
and one spin-$\frac12$ gaugino also in the adjoint representation
of $E_8\times E_8$ or $SO(32)$.

The dilaton $D$ plays a special role in string theory. 
Together with the antisymmetric tensor and the graviton 
it necessarily appears in all (perturbative) string theories.
It is a flat direction of the effective potential  
so that its vacuum expectation value (VEV) 
$\la D\ra$ is a free parameter.
More specifically, this VEV is directly related 
to the string coupling $g_s$ via
\be\label{dilaton}
\la D\ra =\ln g_s \ .
\ee
This can be seen on the one hand from the two-dimensional 
$\sigma$-model approach with an action \cite{GSW,AAT,CT}:
\be\label{sigmamod}
S= S^*(g,b)+\frac{1}{4\pi} \int_\Sigma \sqrt{h} R^{(2)} D(x)\ .
\ee
If one expands the dilaton around its VEV
$D=\la D\ra+\del D$ and uses eq.~(\ref{Euler}) the  action 
$S$ shifts by the constant term $\del S=(2-2n) \la D\ra$. 
This in turn generates a factor of 
$e^{(2 n-2)\la D\ra}$ in the path integral or equivalently
in all scattering amplitudes.
Comparison with eq.~(\ref{sumAs}) then leads to the identification
(\ref{dilaton}).
Alternatively one can  derive (\ref{dilaton}) by explicitly 
calculating appropriate string scattering amplitudes \cite{grosssl}.

\subsection{Low Energy Effective Action}
The space-time spectrum of a string theory
contains a finite number of massless modes, 
which we denote as $L$, and an infinite number of massive modes $H$.
Their mass is an  (integer) multiple  of 
the characteristic mass scale of string theory
$\Mstring$. 
Among the massless modes one always finds a spin-2 object 
which is identified with Einstein's
graviton. This identification relates 
$\Mstring$ to the characteristic scale 
of gravity $M_{\rm Pl}$. More specifically
one finds \cite{DS,kaplu}
\be\label{Mrel}
\Mstring \sim g_{s}^{1-d/2} M_{Pl}
\ee
up to a numerical constant which depends on the
precise conventions chosen.

One is  particularly interested in scattering processes
of massless modes with external momentum $p$
 which is much smaller than  $\Mstring$,
i.e.~one  wants to consider the limit 
$p^2/\Mstring^2 \ll 1$.
The aim is to derive a low energy effective action 
${\cal L}_{\rm eff}(L)$ 
that only depends on the light modes $L$ 
and where all heavy
excitations $H$ have been integrated out. 
This effective action can be reliably computed at 
energy scales far below $\Mstring$.
A systematic procedure for computing ${\cal L}_{\rm eff}(L)$ 
has been developed \cite{GW,grosssl,ltz,kal2} 
and is often referred to as
the $S$-matrix approach. 
One  computes the $S$-matrix
elements 
for a given string vacuum 
as a perturbative power 
series in $g_s$. 
At the lowest order (tree level) an 
$S$-matrix
element typically has a pole in the external momentum 
which corresponds to the exchange of a massless  mode $L$.
The finite part is a power series in 
$p^2/\Mstring^2$ and corresponds to the exchange of the whole
tower of massive $H$-modes. 
${\cal L}_{\rm eff}$ is then constructed to reproduce 
the string $S$-matrix elements in the
limit $p^2/\Mstring^2\ll 1$ with 
$S$-matrix elements constructed entirely
from the effective field theory of the $L$-modes.
In this low energy effective theory the exchange of the $H$-modes in the 
string scattering is replaced by an effective
interaction of the $L$-modes. 
For a four-point amplitude this procedure is 
schematically sketched in figure~4.  
\begin{figure}[htp]
\input{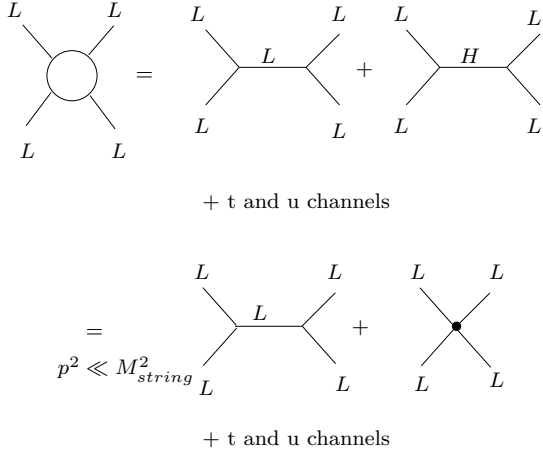}
\caption{The S-matrix approach.}
\end{figure}
The first row denotes the string scattering amplitude and its
separation in a `pole piece' (exchange of a massless mode) 
and the finite piece (exchange of the heavy modes).
The second row indicates ordinary field-theoretical
Feynman diagrams computed from the effective Lagrangian.
The pole piece is reproduced by the same exchange of the massless modes
while the finite part is identified with an effective interaction.
Using this procedure ${\cal L}_{\rm eff}$ can be systematically constructed as a 
power series in both $p^2/\Mstring^2$ and $g_s$.
The power of $p^2$ counts the number of space-time derivatives in 
${\cal L}_{\rm eff}$; at order $(p^2/\Mstring^2)^0$ one finds 
the effective potential while the order $(p^2/\Mstring^2)^2$ 
corresponds to the two-derivative kinetic 
terms.\footnote{Instead of using the S-matrix approach one can  alternatively
construct the effective action by computing the $\beta$-functions
of the two-dimensional $\sigma$-model and interpreting them 
as the equations of motion of string theory.
The effective action is then constructed to reproduce these 
equations of motion 
(for a review see for example \cite{AAT,CT,GSW}).}

\subsection{String Vacua in $d=4$}

For the rest of these lectures we concentrate on string vacua
with four space-time dimensions. 
In this case  the necessary central charges can be red off from table 2.
For the superstring we have 
$(\bar c_{\rm st},c_{\rm st})= (6,6)$ and
$(\bar{c}_{\rm int},c_{\rm int})=(9,9)$
while  for the heterotic string  
$(\bar c_{\rm st},c_{\rm st})= (4,6)$ and
$(\bar{c}_{\rm int},c_{\rm int})=(22,9)$ holds.
The massless spectrum in the space-time or universal sector
can be  obtained by naive dimensional reduction from the $d=10$
massless fields.  
Thus there always are a graviton $g_{mn}, m,n=0,\ldots,3$,
an antisymmetric tensor  $b_{mn}$ and the dilaton $D$.
In $d=4$ the  antisymmetric tensor $b_{mn}$ has one 
physical degree of freedom and  is `dual'
to a Lorentz scalar $a$.\footnote{This duality 
interchanges the Bianchi identity with the field equation
of different tensors of the Lorentz group and has 
no relation with the string duality discussed earlier.}
This duality can be made explicit through the 
field strength $H_{npq}$ of the antisymmetric tensor
\be\label{Haduality}
H^m:=\frac13 \eps^{mnpq}H_{npq}=\eps^{mnpq}\p_n b_{pq}\sim \p^m a(x) .
\ee
$H^m$ is invariant under the local gauge transformation 
$\del b_{mn}=\p_m\xi_n(x) -\p_n\xi_m(x) $ which transmogrifies 
into a continuous Peccei-Quinn (PQ) shift symmetry for the scalar $a(x)$
\be\label{PQ}
a(x)\to a(x)-\frac{\gamma}{4\pi} \ .
\ee
$\gamma$ is an arbitrary real constant and the factor 
$4\pi$ has been introduced
for later convenience. This PQ-symmetry holds 
to all orders in string perturbation theory but as we will see
is generically broken by non-perturbative effects.

The internal sector of
the heterotic string has  central charges
$(\bar{c}_{\rm int},c_{\rm int})=(22,9)$. 
The left moving internal 
$\bar{c}_{\rm int}=22$ CFT
together with the right moving
$\bar{c}_{\rm st}=6$ CFT gives rise to 
(non-Abelian) gauge bosons 
of a gauge group $G$. 
In $d=4$ or equivalently for a
$\bar{c}_{\rm int}=22$ CFT the 
constraint from modular invariance 
is much less stringent as for $d=10$ ($\bar{c}_{\rm int}=16$)
and many gauge groups other than 
$E_8\times E_8$ or $SO(32)$
are allowed.
However, the size of $G$ is not arbitrary but
 bounded by the central charge
$\bar{c}_{\rm int}$\footnote{
The precise bound also depends on the number of
space-time supersymmetries.
For $N=4$ one finds (\ref{rankhetgauge})
while for $N=2$ ($N=1$) one has 
${\rm rank}(G)\le 24$ (${\rm rank}(G)\le 22$).
Furthermore, all of these
bounds only exists in perturbation theory. 
Non-perturbatively the gauge group can be enhanced
beyond the bound imposed by the central charge
\cite{witten}. We briefly return to this point in lecture~4.}
\be\label{rankhetgauge}
 {\rm rank}(G)\le 28 \ .
\ee

The right moving $c_{int}=9$ CFT
can support space-time supercharges 
if it is invariant under additional (global)
worldsheet supersymmetries.\footnote{
Strictly speaking there also is a condition
on the (worldsheet) $U(1)$ charge of the
primary states \cite{bdfm,bd}.
Alternatively, the conditions for space-time
supersymmetry can be stated in terms of
generalized Riemann identities of the partition 
function \cite{lsw}.}
For example, $N=1$ space-time supersymmetry
requires a (global) $(0,2)$ supersymmetry 
of the $c_{int}=9$ CFT  \cite{bdfm}.
In order to obtain $N=2$ space-time supersymmetry 
one has to split the $c_{int}=9$ CFT 
into a free $c_{int}=3$ CFT with $(0,2)$ 
worldsheet supersymmetry and 
a $c_{int}=6$ CFT  with $(0,4)$ supersymmetry
\cite{bd}.
Finally,
a  heterotic vacuum with $N=4$ 
space-time supersymmetry is constructed
by splitting the 
$c_{int}=9$ CFT into three $c_{int}=3$ CFT each with 
$(0,2)$ world-sheet supersymmetry.

The previous discussion related the amount of space-time supersymmetry to 
properties of the internal CFT in particular to the
 amount of
worldsheet supersymmetry. A  subset of these CFT 
can be associated with a compact  
manifold on which the ten-dimensional heterotic string is compactified.
Such compact manifolds have to be  
six-dimensional Ricci-flat
K\"ahler manifolds with a holonomy group contained in 
$SU(3)$; such manifolds are termed Calabi--Yau manifolds
and we summarize some of their properties in appendix B.
$N=1$ is obtained when the holonomy group is exactly $SU(3)$
which corresponds to a Calabi-Yau threefolds $CY_3$.
$N=2$ requires a holonomy group $SU(2)$ corresponding
to the two-dimensional $K3$ surface times a two-torus $T^2$.
Finally a toroidal compactification of the heterotic string
leaves all supercharges intact and thus has $N=4$ supersymmetry.
We summarize the conditions for space-time supersymmetry
in  table~4.

\begin{table}
\begin{center}
\begin{tabular}{|c|c|c|}\hline
\multicolumn{3}{|c|}{\bf Heterotic String}\\\hline
space-time & world-sheet & compact \\
 SUSY      & SUSY        & manifold\\\hline
$N=1$                     & $(0,2)$                        &$CY_3$  \\
$N=2$                     & $(0,4)\oplus (0,2)$            & $K3\times T^2$\\
$N=4$                     & $\! (0,2)\oplus (0,2)\oplus(0,2)\! $ & $T^6$    \\\hline 
\end{tabular}
\end{center}
\caption{Worldsheet and space-time supersymmetry of the heterotic string.}
\end{table}

A similar discussion for the superstring depends on 
the symmetry between left and right moving sectors.
In these lectures we only consider  left-right symmetric
type II vacua and 
without further discussion 
we summarize the relations between 
space-time supersymmetry, worldsheet supersymmetry and
the compactification manifolds  in table~5.

\begin{table}
\begin{center}
\begin{tabular}{|c|c|c|}\hline
\multicolumn{3}{|c|}{{\bf Superstring} (left-right symmetric)}\\\hline
space-time & world-sheet & compact \\
SUSY       & SUSY        & manifold\\\hline
$N=1$                     & -----                      &-----\\
$N=2$                    & $(2,2)$                       & $CY_3$\\
$N=4$                    & $(4,4) \oplus (2,2) $      & $K_3\times T^2$\\\hline 
\end{tabular}
\end{center}
\caption{Worldsheet and space-time supersymmetry of the superstring.} 
\end{table}


\section{$N=1$ Heterotic Vacua in $d=4$}
We start our discussion of the low energy 
effective
action  with the class of four-dimensional heterotic
string vacua whose spectrum and interactions
are  $N=1$ supersymmetric. 
These are also the string vacua
which have mostly been studied so far   
because of  their  phenomenological prospects.
Let us recall  some 
basic facts  of supersymmetry and supergravity 
following the notation and 
conventions of \cite{wessbagger}.

\subsection{$N=1$ Supergravity}

$N=1$
supersymmetry is generated by four 
fermionic charges $Q_{\alpha}$ and 
$Q_{\dot{\alpha}}$, 
which transform as   Weyl spinors 
 of opposite chirality under  the 
Lorentz group.
They obey the  supersymmetry algebra 
\be
 \{Q_{\alpha},\bar{Q}_{\dot{\alpha}}\}
=2\sigma_{\alpha\dot\alpha}^m p_m\ ,
\ee
where $p_m$ is the four-momentum 
and $\sigma^m$ are the Pauli matrices.

There
 are four distinct $N=1$ supersymmetric multiplets
in $d=4$, the gravitational multiplet $E$,
the vector multiplet $V$, the chiral multiplet $\Phi$ 
and finally the linear multiplet $L$. 
The gravitational multiplet  consists of
the graviton $g_{mn}$ and   gravitino 
$\psi_{m\alpha}$;
the vector multiplet features a gauge boson 
$v_m$ and a spin-$\frac{1}{2}$ gaugino $\lambda_{\alpha}$ while the
chiral multiplet contains a complex scalar $\phi$ and a chiral Weyl spinor
$\chi_{\alpha}$. The linear multiplet contains a
real scalar $l$, 
a Weyl fermion $\chi_{\alpha}$ and 
a conserved vector 
$H^m$, which is the field
strength of an antisymmetric tensor 
$H^m=\frac13 \eps^{mnpq}H_{npq}=  \eps^{mnpq}\p_n b_{pq}$.
All four multiplets have two bosonic and two fermionic physical degrees of 
freedom and they  are summarized in  table~6.
\begin{table}
\begin{center}
\bt{|ll|l|}\hline
\multicolumn{2}{|c|}{$N=1$ multiplets} &spin  \\\hline
gravity& $E\sim (g_{mn},\psi_{m\alpha})$&$(2,\frac{3}{2})$\\
vector& $V\sim (v_{m},\lm_{\alpha})$&$(1,\frac{1}{2})$\\
chiral & $\Phi\sim (\chi_{\alpha},\phi)$&$(\frac{1}{2},0)$\\
linear & $L\sim (\chi_{\alpha},H_{m},l)$&$(\frac{1}{2},0)$\\\hline
\et
\end{center}
\caption{$N=1$ multiplets.}
\end{table}

As we already discussed in the last lecture 
an antisymmetric tensor in $d=4$ is dual to a real scalar $a(x)$ 
(c.f.~eq.~(\ref{Haduality})).
In $N=1$ this duality generalizes to a duality between an entire linear and 
a chiral multiplet \cite{fv,BGGM,cfv}. 
In particular the complex scalar $S$ of the dual chiral multiplet
is given by $S=l+i a$  so that
the continuous PQ-symmetry (\ref{PQ}) acts on $S$ according to
\be\label{SPQ}
S\to S-\frac{i\gamma}{4\pi}\ .
\ee
This symmetry  holds to all orders in perturbation theory and 
strongly constrains the possible interactions of the dual chiral multiplet.
We choose to eliminate the linear multiplet from our subsequent
discussions and express all couplings in terms of the dual chiral multiplet.
This simplifies some of the formulas below
but more importantly at the non-perturbative level 
the physics is more easily captured
in the chiral formulation.\footnote{The recent progress
about non-perturbative properties of string vacua
indicate that the appearance of an antisymmetric 
tensor is an artifact of string perturbation theory
and that in a non-perturbative formulation the dilaton
sits in a chiral multiplet \cite{binetruy,dfv,kv,vjs,klm}. 
We return to this aspect in lecture~4.}
However, we should stress that
some of the perturbative properties 
that we will encounter can 
be understood on a more conceptual level 
by using the linear formulation 
\cite{co,dfkz,GT,BGG,ABGG,derqq}.\footnote{
For a discussion of field-theoretical non-perturbative effects
(gaugino condensation) in the linear multiplet
formalism see refs.~\cite{BGT,bdqq,GL}.}

With this in mind let us  recall
the bosonic terms of the 
most general gauge invariant supergravity Lagrangian 
with only chiral and vector multiplets
and no more 
than two derivatives \cite{cfgvp,wessbagger}

\bea\label{genlagr} 
\lefteqn{\cL=-\frac{3}{\kappa^2}\int d^2\theta d^2\bar{\theta} 
E\,  e^{-\frac{1}{3}\kappa^2 K(\Phi,\bar\Phi,V)}}\nonumber\\
& &+  \frac 14 \int d^2\theta \cE \sum_a f_a(\Phi) 
(\cW_{\alpha}\cW^{\alpha})_a+h.c.
\nonumber\\
& &+ \int d^2\theta \cE\, W(\Phi) +h.c.
\nonumber\\
&=&-\rg\Big(\frac{1}{2\kappa^2} R
+G_{I\bar{J}}\, \cD_m\bar{\phi}^{\bar{J}}\cD^m\phi^I
+V(\phi,\bar{\phi})\nonumber\\
& &+\, \sum_a\frac{1}{4 g_a^2}(F_{mn}F^{mn})_a
+\frac{\theta_a}{32 \pi^2}(F\tilde{F})_a\nonumber\\
& & +\quad {\rm fermionic\;\; terms}\Big)
\eea
where $\kappa^2=\frac{8 \pi}{M_{Pl}^2}$;
$E$ is the superdeterminant and $\cE$ the chiral density
(for a precise definition of the superfield action, 
see \cite{wessbagger}).

Supersymmetry imposes constraints on the couplings of 
$\cL$ in eq.~(\ref{genlagr}). 
The metric $G_{I\bar{J}}$ of the manifold spanned by the
complex scalars $\phi^I$
is necessarily a K\"ahler metric and therefore obeys
\be\label{Kmetric}
G_{I\bar{J}}=\frac{\p}{\p\phi^I}\frac{\p}{\p\bar{\phi}^{\bar{J}}}
\, K(\phi,\bar{\phi})\ ,
\ee
where $K(\phi,\bar{\phi})$ is the K\"ahler potential. 
It is an arbitrary real and gauge invariant function of $\phi$
and $\bar{\phi}$.

The gauge group $G$ is in general a product of simple group factors
$G_a$ labelled by an index $a$, i.e.
\be
G=\otimes_a\, G_a\ .
\ee
With each factor $G_a$
there is an associated gauge coupling $g_a$
which can depend on the $\phi^I$.
However, supersymmetry constrains the 
possible functional dependence and demands that 
the (inverse) gauge couplings $g_a^{-2}$ 
are  the real part of a holomorphic function $f_a(\phi)$
called the gauge kinetic functions.
The  imaginary part of the $f_a(\phi)$ 
are  (field-dependent) $\theta$-angles.
One finds
\bea\label{gaugekinf}
g_a^{-2}&=&{\rm Re} f_a(\phi)\ , \nonumber\\
\theta_a&=&-8 \pi^2\, {\rm Im}f_a(\phi)\ .
\eea

The scalar potential $V(\phi,\bar{\phi})$ is also 
determined by a  holomorphic function, the superpotential $W(\phi)$
\be\label{superpot}
V(\phi,\bar{\phi})=e^{\kappa^2 K}\Big(D_I W G^{I\bar{J}}\bar{D}_{\bar{J}}\bar{W}
-3 \kappa^2 |W|^2\Big) ,
\ee
where $D_IW:=\frac{\p W}{\p\phi^I}+\kappa^2\, \frac{\p K}{\p \phi^I}\, W$.

To summarize, 
$\cL$ is completely determined by three  
functions of the chiral multiplets, 
the real K\"ahler potential $K(\phi,\bar\phi)$, 
the holomorphic superpotential $W(\phi)$ and the 
holomorphic gauge kinetic functions $f_a(\phi)$. 

However, there is a certain redundancy in this description.
{}From eq.~(\ref{Kmetric}) we learn that the K\"ahler metric $G_{I\bar{J}}$
is invariant under a harmonic shift of the K\"ahler potential
$K(\phi,\bar{\phi}) \to K(\phi,\bar{\phi}) + F(\phi) +\bar F(\bar{\phi})$.
The entire Lagrangian (\ref{genlagr})
shares this invariance
if the superpotential is simultaneously rescaled while the gauge kinetic
function is kept invariant. 
Altogether $\cL$ is invariant
under 
the replacements 
\footnote{A nowhere vanishing $W$ 
can be completely absorbed by a K\"ahler 
transformation and one defines
$G:=K+\ln|W|^2$ \cite{cfgvp}. However, $W$ usually
does have zeros and it is necessary to keep 
a separate $K$ and $W$.}

\bea\label{Kahler}
K(\phi,\bar{\phi})&\to& K(\phi,\bar{\phi}) + F(\phi) +\bar F(\bar{\phi})\nonumber\\
W(\phi)  &\to& W(\phi)\,  e^{-F(\phi)}\\
f_a (\phi)  &\to& f_a(\phi)\ .
\nonumber
\eea

\subsection{$N=1$ Heterotic String}
Let us now turn to the heterotic string 
and determine some of the generic
properties of $K,W$ and  $f_a$. 
In section~1.2.~we briefly described
a systematic procedure (the S-matrix approach) of how to compute
the effective Lagrangian. Supersymmetry simplifies this project
considerably since it reduced the arbitrary and hence unknown couplings 
to just $K,W$ and  $f_a$.
In section~1.3.~we already discussed the special role played by the
dilaton and its relation to the string coupling.
From eqs.~(\ref{sumAs}),(\ref{dilaton}) one also infers 
that there exists a particular
coordinate frame -- called the string frame -- where
the dilaton multiplies the entire tree level Lagrangian. In this frame  
the bosonic part of the effective action is given by
\bea\label{streffact}
\lefteqn{\cL^{(0)}=\sqrt{\hat{g}} e^{-2 D}\Big\{-\frac{1}{2\kappa^2} \hat{R}
-\frac{1}{4}\sum_a k_a (F_{mn} F^{mn})_a}\nonumber\\
& & \quad
-\, \frac{2}{\kappa^2}\p_mD\p^m D+\frac{1}{16\kappa^2} H_{m}H^{m}\\
& &\quad -\, \tilde{G}_{I\bar{J}}\cD_m\phi^I\cD^m\bar{\phi}^{\bar{J}}
-V(\phi,\bar{\phi})\Big\}\ , \nonumber
\eea
where the $\phi^I$ now denote  all massless scalar fields 
in the string spectrum except the dilaton $D$ and axion $a$.
The constant $k_a$ is a positive integer 
(for non-Abelian gauge groups) 
and is the level of the left moving
$\bar c = 22$ Ka\v{c}-Moody
algebra whose zero modes generate the space-time
gauge bosons.
$H_{m}$ is a modified field strength which also contains Chern--Simons
couplings of the antisymmetric tensor $b_{mn}$ with the gauge fields 
and  gravitons. 
From amplitudes like $\la b_{mn} v_p v_q\ra$ or 
$\la b_{mn} g_{pr} g_{qs}\ra$ one obtains 
\cite{GSW,grosssl,cfgp,cfv,BGG}
\be\label{antisymten}
H^m= \eps^{mnpq}\p_nb_{pq} +\kappa^2 (\om_L^m -\sum_a k_a\, \om_a^m) \ ,
\ee
where $\om_a^m$ is the Yang-Mills Chern-Simons term defined as 
$\om_a^m=\eps^{mnpq}(v_nF_{pq}+\frac{2i}{3}v_nv_pv_q)_a$ 
and $\om_L^m$ is the appropriate Lorentz Chern-Simons term.

In the string frame (\ref{streffact})
 the Einstein term is not canonically normalized and therefore  
the effective string Lagrangian 
cannot yet be
compared with the supergravity 
Lagrangian (\ref{genlagr}).
However, a  Weyl rescaling of the space time
metric 
\be\label{weylresc}
\hat{g}_{mn}=e^{2D}g_{mn}
\ee
in eq.~(\ref{streffact})
results in a  canonical Einstein term.
In addition, one has to 
perform the duality transformation of the antisymmetric tensor\footnote{One
first treats $H^m$ as an 
unconstrained vector and imposes 
the constraint 
$\p_m H^m = \kappa^2(R\tilde R - \sum _a k_a (F \tilde F)_a)$ 
with a Lagrange multiplier $a(x)$.
Then the variation  with respect to $H^m$ implies 
$H_m\sim e^{4 D}\p_m a(x)$.
(For more details see for example refs.~\cite{BGG,ABGG}.)}
and then combine the dilaton $D$ and the axion $a$  
into a complex scalar field 
\be\label{complsc}
S=e^{-2 D}+i a\ .
\ee
After these manipulations 
one arrives at 
\bea\label{strlagr}
\lefteqn{\cL^{(0)}=-\sqrt{g}\Big\{\frac{1}{2 \kappa^2} R\
+\ \tilde{G}_{I\bar{J}}\, \cD_m\phi^I\cD^m\bar{\phi}^{\bar{J}}}\nonumber\\
&+&\! G_{S\bar S}\, \p_m S\, \p^m \bar S +
 \frac{1}{{\rm Re} S}\, V(\phi,\bar{\phi})\\
&+&\! \sum_a \frac{k_a}{4} \left({\rm Re} S\,  (F_{mn}F^{mn})_a
- {\rm Im} S\,  (F\tilde{F})_a\right)\nonumber\Big\},
\eea
where $G_{S\bar{S}}=\frac{1}{\kappa^2 (S+\bar{S})^2}=\p_S\p_{\bar{S}}K$.
Now one  can easily do the comparison with 
 the  supergravity Lagrangian (\ref{genlagr}) and determine
\bea\label{k0w0f0}
K^{(0)}&=&-\kappa^{-2}\ln(S+\bar{S})+\tilde{K}^{(0)}(\phi,\bar{\phi})\nonumber\\
W^{(0)}&=& W(\phi) \;\;{\rm i.e.}\;\; \p_S W^{(0)}=0\\
f_a^{(0)}&=&k_a\,  S\ ,\nonumber
\eea
where $\tilde{G}_{I\bar{J}}=\p_I\p_{\bar{J}}\tilde K$.
Note that the PQ-symmetry  (\ref{SPQ})
shifts the Lagrangian
(\ref{strlagr}) by a total divergence and thus  the perturbative action
is indeed invariant.\footnote{
Here we only considered the bosonic part of the Lagrangian 
but the analysis can be extended to the entire Lagrangian.}

\subsubsection{Perturbative Corrections}
So far the  analysis was confined to the string tree level.
The next step is  to include string loop corrections into the
effective Lagrangian. In section~1.2.~we already determined the relation
between the dilaton and the string coupling constant which organizes
the string loop expansion. In fact, eqs.~(\ref{sumAs}),(\ref{dilaton}) show that the
dilaton dependence  of a given coupling on a 
genus $n$ Riemann surface is fixed.
Therefore all higher loop corrections of the 
functions $K$, $f$ and $W$
are constrained by the following two 
properties:
\begin{itemize}
\item   $e^{2 D} = \frac{2}{S+\bar{S}}$ organizes the string
perturbation theory with large $S$ corresponding to 
weak coupling, 
\item the PQ-symmetry (\ref{SPQ})
is unbroken in perturbation theory.
\end{itemize}

Using these constraints one  expands
$K$, $f_a$ and $W$ in powers of the  dilaton 
in accord with the string loop expansion.
For the K\"ahler potential one finds
\be\label{kexp}
K=K^{(0)}+\sum_{n=1}^{\infty} \frac{\tilde{K}^{(n)}(\phi,\bar{\phi})}
     {(S+\bar{S})^n}\ ,
\ee
where the $\tilde{K}^{(n)}$ are  arbitrary functions of  the scalar fields 
$\phi^I$ but do not depend on the dilaton.
The superpotential $W$ and the gauge 
kinetic function $f_a$ are additionally  constrained by their holomorphy. 
Since $W^{(0)}$ does not depend on the dilaton, the only possible loop corrections
(which are also invariant under the PQ-symmetry)
could look like $W^{(n)}(S,\phi)\sim \frac{W^{(n)}(\phi)}{(S+\bar{S})^n}$ 
for $n>1$, but any such term is
 incompatible with the holomorphy of $W$. 
Therefore $W$ cannot receive 
corrections in string perturbation theory. 
This is  a special case of 
the non-renormalization theorem
of  the superpotential which holds in
any $N=1$ supersymmetric field theory 
\cite{GRS}. The `stringy' proof of this theorem
which we recalled above was first presented
in ref.~\cite{disei}.

The exact same considerations also imply 
a non-renormalization theorem
for the gauge kinetic functions $f_a$ \cite{shifvain,nil}.
One finds that only a dilaton independent one-loop correction 
$f_a^{(1)}(\phi)$ can arise in perturbation 
theory and hence
\be\label{fexp}
f_a=k_a\, S + f_a^{(1)}(\phi)\ .
\ee

\subsubsection{Non-Perturbative Corrections}
Non-perturbative corrections to the couplings 
of the effective Lagrangian generically
break the continuous PQ-symmetry
to its anomaly free discrete subgroup.
Space-time instantons generate a non-trivial topological density
$\frac{1}{32\pi^2}\int F\tilde F$ and therefore break the PQ-symmetry to
\be\label{PQdis}
S\to S-\frac{i n}{4 \pi}\ ,
\ee
where $n$ is an integer and no longer a continuous parameter. 
The holomorphic invariants of (\ref{PQdis})
include  the exponential $e^{-8\pi^2 S}$ and thus 
beyond perturbation theory one expects non-perturbative corrections 
of $W$ and $f$ to have the form:
\bea\label{winstfinst}
W&=&W^{(0)}(\phi)+W^{(NP)}(e^{-8\pi^2 S},\phi)\ ,\nonumber\\
f_a&=&k_a S + f_a^{(1)}(\phi)+f_a^{(NP)}(e^{-8\pi^2 S},\phi)\, .
\eea

To summarize, we learned in this section
that the dilaton dependence of the couplings
$K$, $W$ and $f_a$ is fixed. 
The dependence on all other scalar fields 
$\phi^I$
cannot be determined in general; it
depends on the particular string vacuum under
consideration or in other words the details
of the internal CFT.
Furthermore, the
quantum corrections  of  the holomorphic
$W$ and $f_a$ are strongly constrained
by  non-renormalization theorems.
(The non-renormalization theorems are also discussed
from a slightly different point of view
in lectures by K.~Intriligator.)

\subsection{Supersymmetric Gauge Couplings}
\subsubsection{Preliminaries}
Up to now we denoted all scalars except the dilaton by $\phi^I$. 
Let us introduce a further distinction and 
separate the scalars into matter fields $Q^I$
that are charged under the
gauge group and gauge neutral scalar fields $M^i$, 
called moduli.
The moduli are flat directions of the effective potential in that they
satisfy $\frac{\p V}{\p M^i}=0$ for arbitrary $\la M^i\ra$.
Hence, the VEVs
$\la M^i\ra$ are free parameters of the string vacuum and they can be 
viewed as the coordinates of
a (multi-dimensional) parameter space called the moduli space.
On the other hand 
the vacuum expectation values  of the 
$Q^I$ are fixed by the potential.\footnote{
There also can be gauge neutral singlets which
are not moduli in that there VEV is fixed by the potential.
Such fields are included among the $Q^I$.
Furthermore, there often are also
charged states which are
flat directions of $V$. Their VEVs break the gauge
group and give the associated gauge bosons a mass.
In the effective field theory description
one has a choice to either integrate out these
massive states  
along with the whole tower of
heavy string modes or leave them in the low energy
effective action. The latter choice is appropriate
when the masses are small and well below the cutoff
of the effective theory. In this case we include
these flat directions among the $Q^I$.
The first choice is sensible whenever the
masses are close to the string scale $\Mstring$. Once they are integrated out the gauge group
is reduced and only gauge neutral states 
are left over.
The important point is that flat directions
that are charged change the low energy spectrum while 
gauge neutral flat directions are spectrum 
preserving.
For a more detailed discussion of this distinction
see for example \cite{vadjan1}.
 }
One conveniently chooses $\la Q^I\ra=0$ and
expands all couplings around this expectation value. In particular we  need 
\bea\label{kwfexp}
\tilde{K}(\phi,\bar{\phi})&=&\kappa^{-2}\ \hat{K}(M,\bar{M})\nonumber\\
& &+\  Z_{\bar{I}J}(M,\bar{M})\ 
  \bar{Q}^{\bar I} e^{2 V} Q^J
 +\ldots \nonumber\\
f^{(1)}(\phi)&=& f_a(M)\ +\ \ldots\ ,
\eea
where the ellipsis stand for higher order terms 
that are irrelevant for our purpose.
The couplings 
$\hat{K}(M,\bar{M})$, $Z_{\bar{I}J}(M,\bar{M})$ 
and $f_a(M)$ do not depend 
on the dilaton but only on the moduli;
in general this  functional dependence 
cannot be further specified.

\subsubsection{Field Theory Considerations}
Any effective field theory has two distinct kinds of gauge couplings, 
a momentum dependent (running) effective gauge coupling $g_a(p^2)$, 
and a  Wilsonian gauge coupling.
Shifman and Vainshtein 
\cite{shifvain} stressed the importance of this distinction for 
supersymmetric field theories.
It arises from the fact that the
Wilsonian gauge coupling is the real part of a holomorphic function
${\rm Re} f_a$ which is not renormalized beyond one-loop.
This Wilsonian coupling is the gauge coupling
of a Wilsonian effective action which is defined by only integrating out
the heavy and high frequency modes with momenta  above a given
cutoff scale. 
By construction such a Wilsonian Lagrangian is local and its 
couplings obey the analytic and renormalization properties 
discussed in the previous section.\footnote{For further
discussion about the Wilsonian Lagrangian see for
example refs.~\cite{vadjan1,DShir}.}
On the other hand $g_a(p^2)$ are the couplings in the 
one-particle irreducible generating  
functional which includes momenta at all scales;
it is related to physical quantities such as scattering
amplitudes.
At the tree level the two couplings coincide and we have 
$(g_a^{(0)})^{-2} =  {\rm Re} f_a^{(0)}$. However, at the one-loop level
they start to disagree and for any 
locally supersymmetric field theory
one finds  instead
\cite{shifvain,louis,co,dfkz,GT,BGG,vadjan1,vadjan2}
\bea\label{ftg}
\lefteqn{g_a^{-2}(p^2)={\rm Re} f_a
+\frac{b_a}{16 \pi^2} \ln\frac{\La^2}{p^2}
+\frac{c_a}{16 \pi^2} \hat K^{(0)} }\\
&+& \frac{T(G_a)}{8 \pi^2}\ln g^{(0)-2}_a
-\sum_r\frac{T_a(r)}{8 \pi^2}\ln\det Z_{r}^{(0)}
\ , \nonumber
\eea
where $r$ runs over the representation of the gauge group and $\La$ is 
a (moduli independent) UV cutoff of the regularized supersymmetric 
quantum field theory
which  is naturally chosen  as the Planck mass $\La=M_{Pl}$.\footnote{
The important point here is that $M_{Pl}$ is the field independent
mass scale of the supergravity Lagrangian (\ref{genlagr}).}
$\hat K^{(0)}$ and $Z_{r}^{(0)}$ are the tree level couplings of the 
light (or massless) modes and $b_a$ is the one-loop coefficient of
their $\beta$-function.\footnote{More precisely,
$Z_{r}$ is the block of $Z_{\bar{I}J}$ 
corresponding to the 
matter  multiplets in representation $r$.}
We also abbreviated 
\bea
T_a(r)\, \delta^{(a)(b)}&=&\tr_a T^{(a)}T^{(b)}\nonumber \\
T(G_a)&=&T_a({\rm adj.\ of}\ G_a)\nonumber\\
b_a&=&\sum_r n_r T_a(r)-3\, T(G_a)\\
c_a&=&\sum_r n_r T_a(r)-T(G_a)\ ,\nonumber
\eea
where $T^{(a)}$ are the generators of the gauge group and
$n_r$ denotes the number of matter multiplets in representation $r$.

There are several points about
eqs.~(\ref{ftg})  which need to 
be stressed:
\begin{itemize}
\item The effective gauge couplings are not 
      harmonic functions of the moduli, 
      that is $\p_i \p_{\bar{j}} g_a^{-2}\neq 0$. This failure of
      harmonicity is known
      as the holomorphic or K\"ahler anomaly. 
      It implies that $g_a^{-2}$ is not the real part of 
      a holomorphic function;
      $ g_a^{-2}={\rm Re}f_a$ only holds at the tree level
      but not when higher loop corrections are included.\footnote{A similar situation
      has been found for the superpotential \cite{west,jjw,djj,vadjan1,NS,PR}. 
      Beyond the tree level 
      it is necessary to make a distinction between the effective Yukawa couplings
      and the holomorphic Wilsonian parameters of the superpotential.
      They coincide at the tree level but not beyond when massless modes
      are in the spectrum. The holomorphic superpotential $W$ is by construction
      a Wilsonian quantity and obeys all the non-renormalization theorems
      of the previous section.}

\item The non-harmonic differences $g_a^{-2}-{\rm Re} f_a$ 
      only depend on the 
      massless  modes and their couplings. 
      Therefore they can be computed
      entirely in the low energy effective field theory 
      without any additional knowledge
      about the underlying fundamental theory.
\item The Wilsonian couplings $f_a$ are always  holomorphic and 
      only corrected at the one-loop level and non-perturbatively.
      These quantum corrections are
      induced by the heavy modes 
      of the underlying fundamental theory.
      
\item The axionic couplings still obey
      $\p_{M^j}\theta_a = 8i\pi^2  \p_{M^j} g_a^{-2}$ but 
      $\p_{M^j}\theta_a$ is no longer integrable whenever
      $g_a^{-2}$ is non-holomorphic \cite{kal}.
      The $\theta$-angles are not well defined when massless
      modes are present.
\end{itemize}

The effective gauge couplings $g_a(p^2)$ are physical quantities 
and hence must be invariant under all exact
symmetries of the theory. This is certainly assured at the classical
level but from eqs.~(\ref{ftg}) we learn that such an invariance
is potentially  spoiled at the quantum level.
For example a holomorphic coordinate transformation
\bea\label{ztrans}
Q^I&\to& \Sigma^{I}{}_{J}(M)\,  Q^J\\
Z_{\bar{J}I}&\to&\bar{\Sigma}^{-1}_{\bar{J}}{}^{\bar{K}}\;Z_{\bar{K}L}\Sigma^{-1}_I{}^{L}
\nonumber
\eea
or a K\"ahler transformation
\bea\label{Ktrans}
\hat{K}&\to&\hat{K}+F(M)+\bar{F}(\bar{M})\nonumber\\
W&\to& W e^{-F(M)}
\eea
do not leave $g_a(p^2)$ of eqs.~(\ref{ftg}) invariant.
This is a one-loop violation of two  classical invariances
and therefore they are often referred to as the 
$\sigma$-model and  K\"ahler anomaly.
However,  these anomalies are harmonic functions of the moduli and therefore
can be cancelled by assigning a new one-loop transformation law to 
the gauge kinetic functions $f_a$.
Cancellation of the combined anomalies requires 
the Wilsonian couplings to transform as:
\be\label{ftra}
f_a\to f_a-\frac{c_a}{8 \pi^2}F-\sum_r\frac{T_a(r)}{4 \pi^2}\ln\det \Sigma^{(r)}.
\ee
In that sense  $f_a(M)$ can be viewed as  local counterterms
cancelling potential 
$\sigma$-model and  K\"ahler anomalies  
\cite{louis,co,dfkz,GT,BGG,vadjan1,vadjan2}.

Finally, based on ref.~\cite{NSVZ}
one can write down an all-loop generalization
of eqs.~(\ref{ftg}) \cite{dfkz,vadjan1}
\bea\label{ftgall}
\lefteqn{g_a^{-2}(p^2)={\rm Re} f_a
+\frac{b_a}{16 \pi^2} \ln\frac{M_{\rm Pl}^2}{p^2}
+\frac{c_a}{16 \pi^2} \hat K }\\
&+& \frac{T(G_a)}{8 \pi^2}\ln g^{-2}_a(p)
-\sum_r\frac{T_a(r)}{8 \pi^2}\ln\det Z_{r}(p)
\ , \nonumber
\eea
where now $\hat K, Z_{r}(p)$ are the full loop-corrected
couplings. Eqs.~(\ref{ftgall}) are the solutions of the 
all-loop $\beta$-functions proposed in \cite{NSVZ}
\be
\beta_a (g_a)\equiv p\, \frac{dg_a}{dp} = \frac{g_a^3}{16\pi^2}
\frac{b_a + \sum_r T_a(r)\,  \gamma_r}{1-\frac{T(G_a)}{8\pi^2} g^2}\ ,
\ee
where $\gamma_r := p\frac{d}{dp} \ln\det Z_{r}(p)$.

\subsubsection{Gauge Couplings in String Theory}
The next step is to apply the `lessons'
of the  previous section to 
string theory and explicitly compute the 
low energy gauge couplings. 
First of all, the dilaton dependence of
$g_a$ can be determined by
inserting eqs.~(\ref{Mrel}),(\ref{k0w0f0}),(\ref{fexp})
into (\ref{ftg}) which results in
\be\label{stg}
g_a^{-2} = k_a {\rm Re}S+\frac{b_a}{16 \pi^2}
\ln\frac{\Mstring^2}{p^2} 
  +\Del_a(M,\bar{M})
\ee
where
\bea\label{threshold}
\lefteqn{\Del_a(M,\bar{M})=\R f_a^{(1)}(M)
+\frac{c_a}{16\pi^2} \hat{K}^{(0)}(M,\bar{M})}\nonumber\\
& & - \sum_r \frac{T_a(r)}{8\pi^2}
\ln\det Z_r^{(0)}(M,\bar{M})\ ,
\eea
and
\be\label{Msplrel}
\Mstring^2 \sim \frac{M_{\rm Pl}^2}{S+\bar S}\ .
\ee
The precise numerical coefficient which relates
the scales $\Mstring$ and $M_{\rm Pl}$
is  a matter of convention 
(see for example \cite{kapl})
and the discussion here neglects all 
field independent corrections of $g_a$.
From eqs.~(\ref{stg})--(\ref{Msplrel}) 
we learn that 
the entire one-loop dilaton dependence 
of eq.~(\ref{ftg})
conspires into a  $\ln (S+\bar{S})$ term  with 
the one-loop beta function 
as a coefficient. This amounts to nothing but  
a universal change of the coupling's unification scale.
In string theory the natural starting point for the renormalization of the 
effective gauge couplings is not $M_{Pl}$ but rather
the string scale $\Mstring$. 
That is, string theory and field theory 
naturally choose different cutoffs.
Furthermore, 
the threshold corrections 
$\Del_a$ are  entirely independent of the dilaton 
($\p_S \Del_a=0$) (this fact 
can also be derived  directly from 
$\cL_{\rm eff}$ in the string frame
eq.~(\ref{streffact}) \cite{ms2}).

The moduli dependence of 
$\Del_a$ can be computed in two different ways.
First of all one can explicitly evaluate the
one-loop string diagram (fig.~5) with two gauge boson 
vertex operators  $V_{v_m}$  
as the external legs 
\cite{kapl,kal,agntfirst,agn,ms3,KK}. 
In order to do this computation one needs to know
the exact massive spectrum which runs in the loop
or in other words the 
CFT correlation functions  have to be known.

\begin{figure}[htp]
\input{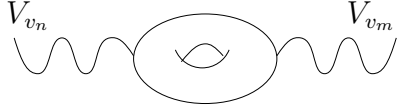}
\caption{Two-point genus one string amplitude with two gauge bosons.}
\end{figure}

A second possibility to determine $\Del_a$ 
makes use of the  constraints implied by the 
holomorphy of  $f_a$,
its quantum symmetries and singularity structure.
This second method will be discussed 
in some detail in the next section.
It has the advantage that it can be used
even if the CFT correlation functions are unknown.
However, in some case the constraints on
$f_a$ are not strong enough to determine
it uniquely but only up to a small number 
of undetermined numerical coefficients
\cite{ms3}.

\subsubsection{Orbifold compactifications}

As an example we consider a specific class of 
orbifold compactification of the
ten-dimensional heterotic string \cite{dhvw,kal}.
An orbifold is constructed from a smooth  toroidal compactification
by dividing the six-torus $T^6$ by a non-freely acting discrete group.
In order to preserve $N=1$ supersymmetry  this discrete group 
should be a subgroup of $SU(3)$  and an isometry of $T^6$. 

Here we focus on a particular subclass of such orbifold compactifications
(factorizable orbifolds)
where the geometrical moduli of 
(at least) one two-torus $T^2$ of $T^6$ 
are left unconstrained in the spectrum.
An example of a factorizable orbifold is 
the ${\bf Z}_4$ orbifold with the generator 
$\theta=(i,i,-1)$ acting on the three complex 
coordinates of $T^6=T^2\times T^2\times T^2$. 
The moduli of the third $T^2$ are 
\bea\label{TUdef}
T&=&2(\sqrt{g}-i b)\ ,\nonumber\\
U&=&\frac{1}{g_{11}}(\sqrt{g}-i g_{12}) \ ,
\eea
where $g_{ij}$ is the background metric on $T^2$,
$g={\rm det}(g_{ij})$ and $b_{ij}=b \eps_{ij}$ 
is the antisymmetric tensor.\footnote{For a precise definition 
of factorizable orbifolds see ref.~\cite{vadjan2}.}

The tree level couplings of these toroidal moduli
are given by \cite{w,clo,fp,kal2,il,vadjan2}
\bea\label{KTU}
\hat{K}^{(0)}&=&-\ln(T+\bar{T})-\ln(U+\bar{U})\, ,\nonumber \\
Z_{I\bar{J}}^{(0)}&=&\del_{I\bar{J}} \,
(T+\bar{T})^{-q_I} (U+\bar{U})^{-q_I} \ ,
\eea
where the $q_I$ are rational numbers depending on the twist sector 
of the orbifold; they can be found 
for example in ref.~\cite{vadjan2}.

Factorizable 
orbifold compactifications always have an 
$SL(2,{\bf Z})_T \times SL(2,{\bf Z})_U $ quantum symmetry \cite{DVV,gpr}.
That is, the partition function as well as all
correlation functions respect this symmetry to all orders
in string perturbation theory.
Such quantum symmetries commonly arise in string vacua and they
are termed $T$-dualities.
For the particular case at hand the 
$SL(2,{\bf Z})_T$ acts  on the  toroidal moduli according to:
\be\label{SL}
T\to\frac{a T-i b}{i c T+d}\ , \quad U\to U\ ,
\ee
where $a,b,c,d\in{\bf Z}$ and $ad-bc=1$. 
The $SL(2,{\bf Z})_U$
has a similar action with $T$ and $U$ interchanged.
(Further details of the modular group $SL(2,\bf{Z})$ are collected in  appendix A.)
Under the transformation (\ref{SL})
the K\"ahler potential of eq.~(\ref{KTU}) undergoes a K\"ahler shift
of the form (\ref{Ktrans})  with
\be\label{f}
F(T)=\ln(i cT+d)\ .
\ee
(The same shift is found for $SL(2,{\bf Z})_U$ 
but with  $T$ replaced by $U$.)\footnote{
Note that although the tree level K\"ahler potential is corrected 
in string perturbation theory the holomorphic function $F(T)$ 
has to be exact to all orders of perturbation theory.
This follows from the fact that the superpotential $W$ 
is protected from any perturbative renormalization.}
From eq.~(\ref{KTU}) one also infers that 
$Z^{(0)}_{I\bar{J}}$ transforms 
according to eq.~(\ref{ztrans}) with
\be
\Sigma^{I}{}_{J}=\del^{I}{}_{J}\, (i c T+d)^{-q_I}\ .
\ee
Inserting (\ref{f}) and (\theequation) 
into (\ref{ftra}) and 
using the fact that the dilaton $S$ 
can be chosen  modular invariant 
one finds that the one-loop corrections $f_a^{(1)} (T,U)$
have to transform  like
\be\label{fmtra}
f_a^{(1)} (T,U) \to f_a^{(1)}(T,U) - \frac{\alpha_a}{8\pi^2}\ln (i c T+d)
\ee
where 
\be
\alpha_a=\sum_I T_a(Q^I)(1-2 q_I)-T_a(G)\ .
\ee
The logarithm of the 
Dedekind $\eta$-function (defined in appendix A)
has precisely  
the transformation properties 
needed to satisfy  (\ref{fmtra}).
More specifically one has  
\be
\eta^2\to\epsilon\eta^2 (i c T+d)\ ,
\ee
where $\epsilon^{12}=1$.
Hence one infers
\bea\label{feta}
f_a^{(1)} &=& -\  \frac{\alpha_a}{8\pi^2}
\ln [\eta^2(iT)\eta^2(iU)] \nonumber\\
& &  +\  P_a[j(iT),j(iU)]\ ,
\eea
where 
$P_a$ are modular invariant holomorphic functions of the moduli and
thus can only depend on $T$ and $U$ through the modular invariant 
holomorphic $j$-function defined in appendix A. 

In order to determine  $P_a$ 
we also need to know the  singularities of $f_a$.
Such singularities appear at points in the moduli
space where some otherwise heavy states become massless.
For factorizable  orbifolds 
there are indeed additional 
massless states at 
$T=U$ (mod $SL(2,{\bf Z})$)  but they are always 
neutral with respect to the low energy gauge group. 
Thus the gauge couplings do not develop a 
singularity  at $T=U$ 
and consequently the  $f_a$ cannot
have any singularities at 
finite $T$ or  $U$ in the moduli space. 
On the other hand, in the limit 
$\R\, T, \R\, U \to \infty$ the theory decompactifies and 
a whole tower of Kaluza--Klein states turns massless.
However, the corresponding singularity in the gauge couplings
is constrained by the fact that there has to exist a region 
in the moduli space where  the theory stays  perturbative
in the decompactification limit. 
This region is characterized by the requirement that both 
the four-dimensional gauge coupling $g_4$ and the six-dimensional
gauge coupling $g_6$ are small.
In the large radius limit the two coupling are 
related by
\be
g_4^{-2} \sim r^2 g_6^{-2}\ ,
\ee 
where $r=\sqrt{\R\, T}$ ({\it c.f.}~(\ref{TUdef}))
is  the radius of the torus. 
Therefore, $g_6$ stays small for 
$g_4^2\cdot \R\, T$ fixed and small.
$g_4$ stays small for $\R\, S$ large and $\R\, S \gg \Delta$
where this last condition merely states that the one-loop
correction is small compared to the tree level value.
This  constrains $\Delta$ or equivalently $f_a(T,U)$
to grow at most linearly in $T$ or
\be\label{largef}
\lim_{\R\, T \to \infty} f_a \to  T\ .
\ee
(The same argument also holds for $U$ since the large $U$ limit 
 is $SL(2,{\bf Z})$ equivalent to the 
decompactification limit.)
From its definitions in appendix~A we learn that 
$j$ and $\eta$ are non-singular inside the fundamental domain
and in the limit $T\to \infty$ their  asymptotic behaviour is
\be\label{largeT}
\ln (\eta^2)\to -\frac{\pi}{12}T\ ,\quad
j\to e^{2\pi T}\ .
\ee
We see that the first term in 
eq.~(\ref{feta}) does satisfy (\ref{largef}). 
The modular invariant $P_a[j(iT), j(iU)]$ 
also have to satisfy (\ref{largef}) but as
we argued earlier they also have to be
finite inside the modular domain. The unique  solution to these
three constraints is  $P_a = {\rm const.}$
\cite{vadjan2}.

To summarize, from the knowledge of the 
tree level couplings $\hat K^{(0)}$ and $Z^{(0)}$ 
and the fact that the modular symmetries
$SL(2,{\bf Z})_T \times SL(2,{\bf Z})_U$ hold at any order in perturbation theory
one uniquely determines
\be
f_a(S,T,U)=k_a S-\frac{\alpha_a}{8\pi^2}
\ln[\eta^2(i T)\eta^2(i U)]
\ee
up to (gauge group dependent)  constants.
Inserted into eq.~(\ref{threshold})
one finds
\be\label{exdel}
\Delta_a=-\frac{\alpha_a}{16\pi^2}
 \ln \Big[|\eta(iT)|^4 \R T\;|\eta(iU)|^4 \R U\Big] .
\ee

The exact same method presented here can also
be applied to Calabi--Yau vacua with a low number
of moduli and in many  cases 
one is able to determine $f_a$ (up to a universal
gauge group independent factor)
\cite{bcov,cofkm,cfkm,hkty2,BKK,vadjan2}.
However, in some cases (for example for
non-factorizable orbifolds)
the constraints on $f_a$ are not strong
enough to determine it uniquely \cite{ms3}.
Instead one is left with a finite number
of undetermined coefficients.

%
\section{$N=2$ String Vacua}
In this section  we study the holomorphic couplings for
heterotic and type II string vacua which have an extended 
$N=2$ space-time supersymmetry. 
In part this is motivated by the work of 
Seiberg and Witten who determined  the exact non-perturbative low energy effective theory
of an $N=2$ supersymmetric Yang--Mills theory \cite{sw}.
In this case the entire low energy effective action is encoded
in terms of a holomorphic prepotential $\cF$.
(The analysis of Seiberg and Witten and its generalizations is nicely
reviewed in the lectures of W.~Lerche.)
It was of immediate interest to also discover  the Seiberg--Witten theory
as the low energy limit of an appropriate string vacuum. 
(This aspect
is reviewed 
in the lectures of S.~Kachru.)

In this lecture we recall the structure of $N=2$ supergravity
with special emphasis on the holomorphic prepotential
and focus on perturbative properties of heterotic  as well as type II 
$N=2$ string vacua.

%
\subsection{$N=2$ Supergravity}
$N=2$ extended space-time supersymmetry is generated by 
two  (Weyl-) supercharges $Q_{\alpha}^A$ ($A=1,2$)  which obey
\bea
\{Q_{\alpha}^A,\bar{Q}_{\dot{\beta}}^{B}\}&=&
    2 \del^{AB} \sigma^{m}_{\alpha\dot{\beta}}p_m\nonumber\\
\{Q_{\alpha}^A,Q_{\beta}^B\}&=&\epsilon_{\alpha\beta} \epsilon^{AB} \cZ\ ,
\eea
where  $\cZ$ is the central charge of the algebra.

The $N=2$ gravitational multiplet $E$ contains the graviton, two gravitini
$\psi_{m\alpha}^A$ and an Abelian vector boson $\gamma_m$ called
the graviphoton.  In terms of $N=1$ multiplets it is the sum of the $N=1$
gravitational multiplet and a gravitino multiplet $\Psi$ which contains
a gravitino and an Abelian vector.\footnote{This multiplet was constructed in ref.~\cite{gg};
it cannot be consistently coupled in $N=1$ supersymmetry.}
An $N=2$ vector multiplet contains a vector,  two gaugini 
$\lambda_{\alpha}^A$ and a complex scalar $\phi$;
it consists of an $N=1$ vector multiplet $V$ and a chiral multiplet $\Phi$.
Matter fields arise from $N=2$ hypermultiplets $H$
which contain two Weyl fermions $\chi_{\alpha}^A$
and four real scalars $q^{AB}$; they are the sum of two
$N=1$ chiral multiplets.
There are three further multiplets which all contain an antisymmetric tensor
field and therefore also will accommodate the dilaton of string theory.
First there is a vector-tensor multiplet $VT$ which contains an Abelian vector,
two Weyl fermions, the field strength of an antisymmetric tensor
and a real scalar; it consists of an $N=1$ vector multiplet 
and a linear multiplet $L$
\cite{ssw,jl,cwfkst,how}.
The tensor multiplet $T$ contains two Weyl fermions, 
the field strength of an antisymmetric tensor
and three  real scalars; it consists of an $N=1$ chiral multiplet 
and a linear multiplet \cite{dwvh}.
Finally, the double tensor multiplet $\Pi$ 
contains  two Weyl fermions, two  field strengths of antisymmetric tensors
and two  real scalars; 
it consists of two  linear multiplets.\footnote{
Dimensional reduction of type IIB string theory 
implies the existence
of this multiplet ({\it c.f.}~table 3)
but as far as we know
it has not been explicitly constructed yet.
Therefore, we leave it as an exercise for the reader.}
All multiplets have four on-shell bosonic and fermionic degrees of freedom
and we summarize their field content and spins  in table~7. 

\begin{table}
\begin{center}\label{n2multiplets}
\begin{tabular}{|l|l|l|}\hline
$N=2$ multiplets &           $N=1$                          & spin           \\ \hline
$E\sim (g_{mn},\psi_{m\alpha}^A,\gamma_m) $ & E  $\oplus \Psi$ &  $(2,\frac32,1)$ \\
 $V\sim (v_m,\lambda_{\alpha}^A,\phi)$& $V\oplus \Phi$& $(1,\frac12,0)$ \\
 $H\sim (\chi_{\alpha}^A,q^{AB})$& $\Phi\oplus \Phi$& $(\frac12,0)$ \\
 $VT\sim(v_m,H_m,\chi_{\alpha}^A,l)$ & $V\oplus L$& $(1,\frac12,0)$\\
$T\sim(H_m,\chi_{\alpha}^A,\phi,l)$& $\Phi\oplus L$& $(\frac12,0)$ \\
 $\Pi\sim (H_m,H'_m,\chi_{\alpha}^A,l,l')$ &$L\oplus L$& $(\frac12,0)$\\\hline
\end{tabular}
\end{center}
\caption{Multiplets of $N=2$ supergravity in $d=4$.}
\end{table}

As we discussed in the previous lectures  an 
antisymmetric tensor is dual  to a pseudoscalar in $d=4$ and this 
translates into the following dualities among $N=2$ multiplets
\be
VT\sim V, \quad
T \sim H, \quad
\Pi\sim H\ .
\ee
The dilaton which arises from  the universal sector of 
the CFT is always accompanied by an antisymmetric tensor.
More specifically, 
for any heterotic vacuum the dilaton is a member 
of a vector-tensor multiplet while 
the dilaton in type IIA (type IIB) vacua resides 
in a tensor (double-tensor)
multiplet (table~8). This can be derived by dimensionally
reducing the ten-dimensional string theories
summarized in table~3.

\begin{table}
\begin{center}\label{dilatonmultiplets}
\begin{tabular}{|l|c|}\hline
string theory &dilaton multiplet \\\hline
heterotic&$VT\sim V$\\
type IIA& $T\sim H$\\
type IIB & $\Pi\sim H$\\\hline
\end{tabular}
\end{center}
\caption{Dilaton multiplet.}
\end{table}
As in lecture~2  we always choose to discuss the low energy
effective theory
in terms of the dual multiplets, 
that is we express the action solely in terms 
of the gravitational multiplet, the vector  
and hypermultiplets.

$N=2$ supergravity severely constrains the interactions 
among these multiplets \cite{bw,wp,wlp,cdaf,daff,abc,abcm}.
In particular,
the complex scalars $\phi$ of the vector multiplets
are coordinates on a special K\"ahler manifold $\cM_V$ while  the
real scalars $q^{AB}$ of the hypermultiplets are coordinates 
on a quaternionic manifold $\cM_H$.
Locally the two spaces form a direct product \cite{wlp}, 
i.e.
\be\label{modulispace}
\cM=\cM_V\otimes \cM_H \ .
\ee
For their respective dimensions 
we abbreviate
${\rm dim}(\cM_V)\equiv n_V$ and 
${\rm dim}(\cM_{H})\equiv n_H$.

%
\subsubsection{Special K\"ahler manifolds}
A special K\"ahler manifold is  a K\"ahler manifold  whose geometry 
obeys  an additional constraint \cite{wp}.\footnote{For
a review of special geometry, see for example ref.~\cite{VPreview}.}
One way to express this constraint  is the statement that the 
 K\"ahler potential $K$ is not an arbitrary real function 
(as it was in $N=1$ supergravity) but 
determined in terms of a holomorphic prepotential $F$
according to\footnote{Alternative definitions can be found in
refs.~\cite{str,cdaf,CDFLL} and their equivalence 
is discussed in ref.~\cite{crwp}.}
\be\label{Kspecial}
K=-\ln\Big(i \bar{Z}^{\Sigma} (\bar\phi) F_{\Sigma}(Z)
- i Z^{\Sigma}  (\phi)\bar{F}_{\Sigma}(\bar{Z})\Big) \ .
\ee
The $Z^{\Sigma}, \Sigma=0,\ldots, n_V$ are $(n_V+1)$
holomorphic functions of the $n_V$ complex scalar fields 
$\phi^I, I=1,\ldots,n_V$ which reside 
in the vector multiplets.
$F_{\Sigma}$ abbreviates the derivative, i.e.
$F_{\Sigma}\equiv \frac{\p F(Z)}{\p Z^{\Sigma}} $
and 
$F(Z)$ is  a homogeneous function of $Z^{\Sigma}$ of degree $2$:
\be
Z^{\Sigma} F_{\Sigma}=2 F\ .
\ee
The  K\"ahler metric $G_{IJ}$ is obtained from eq.~(\ref{Kmetric})
with the   K\"ahler potential (\ref{Kspecial}) inserted.

The above description is again somewhat redundant. The holomorphic
$Z^{\Sigma}$ can be eliminated by an appropriate choice of
coordinates and a choice of the K\"ahler gauge. 
The corresponding coordinates are called 
 special coordinates and  are defined by:
\be
\phi^I=\frac{Z^I}{Z^0}\ .
\ee
In these special coordinates 
the K\"ahler potential  (\ref{Kspecial}) reads
\be\label{Kspecial2}
K=-\ln\Big(2 (\cF+\bar{\cF})-(\phi^I-\bar{\phi}^I)(\cF_I-\bar{\cF}_I)\Big) ,
\ee
where $\cF(\phi)$ is an arbitrary holomorphic function of 
$\phi^I$ related to $F(Z)$ via
$F(Z)=-i (Z^0)^2 \cF(\phi)$.

A lengthy but straightforward computation shows that using 
(\ref{Kspecial}) the Riemann curvature tensor
of special K\"ahler manifolds satisfies \cite{ckpdfwg}
\bea\label{Rspecial}
R_{I\bar{J}K\bar{L}}&=&G_{I\bar{J}}G_{K\bar{L}}+G_{I\bar{L}} G_{K\bar{J}}\nonumber\\
& & - e^{2 K} W_{IKM} G^{M\bar{M}} \bar{W}_{\bar{M}\bar{J}\bar{L}}\ ,
\eea
where $W_{IKM}\equiv\p_I\p_K\p_M \cF$.\footnote{The
$W_{IKM}$ are sometimes  referred to as
the  Yukawa couplings since for a particular class
of heterotic $N=1$ vacua (vacua which have a
global $(2,2)$ worldsheet supersymmetry)
they correspond to space-time Yukawa couplings 
\cite{kal2}.}

$N=2$ supergravity not only constrains the K\"ahler
manifold of the vector multiplets but also relates the gauge couplings 
to the holomorphic prepotential. More specifically,
the gauge kinetic terms are 
\be\label{Lgauge}
\cL=-\frac{1}{4} g^{-2}_{\Sigma\La}\, F_{mn}^{\Sigma} F^{mn\La}
          -\frac{\theta_{\Sigma\La}}{32\pi^2}\, F^{\Sigma}\tilde{F}^{\La}
+\ldots ,
\ee
where $F_{mn}^{0}$ is the field strength of the graviphoton and 
\bea\label{gdef}
g^{-2}_{\Sigma\La}&=&\frac{i}{4} (\cN_{\Sigma\La}-\bar{\cN}_{\Sigma\La})\ ,
\nonumber \\
\theta_{\Sigma\La}&=&2 \pi^2 (\cN_{\Sigma\La}+\bar{\cN}_{\Sigma\La}) \ .
\eea
$\cN_{\Sigma\La}$ is defined by:
\be\label{Ndef}
\cN_{\Sigma\La}=\bar{F}_{\Sigma\La}+2 i \ \frac{\I F_{\Sigma\Omega}\;\I F_{\La\Xi}\;Z^{\Omega} Z^{\Xi}}
{\I F_{\Omega\Xi}\, Z^{\Omega} Z^{\Xi}} \ .
\ee
Due to the second term of $\cN_{\Sigma\La}$ 
the gauge couplings are generically non-harmonic  functions of the moduli.
This is different from the situation in $N=1$ 
since any Abelian factor in the gauge group
can have a non-trivial mixing with the graviphoton
and this is origin of the 
non-harmonicity of the (Wilsonian) gauge couplings in $N=2$
supergravity.

To summarize, both the  K\"ahler potential and the Wilsonian gauge couplings 
of the vector multiplets in an $N=2$ effective Lagrangian are determined by a single
holomorphic function  of the scalar fields $\phi$
-- the prepotential $\cF(\phi)$.
The Wilsonian gauge couplings are non-harmonic already at the tree level.

It is again convenient to make a distinction between 
gauge neutral moduli scalars $M^i$ 
which are members
of Abelian vector multiplets 
and  charged scalars  $Q^I$ which
arise from non-Abelian vector multiplets.\footnote{As we
discussed in the previous section this distinction
is somewhat ambiguous and involves a choice
of the low energy degrees of freedom.
In $N=2$ the scalars in the Cartan subalgebra
of a non-Abelian gauge factor $G_a$ always are
flat directions of the effective potential.
Thus at a generic point in their moduli space
$G_a$ is broken to its maximal Abelian subgroup.}
That is, we split the scalars $\phi^I$ according to 
$\phi^I = (M^i, Q^I)$\footnote{As before we use the index $I$ in two different
ways and hope the reader will not be confused by this notation.}
and expand $\cF$ and $K$  as a truncated 
power series
around $\la Q^I\ra=0$ exactly as in
$N=1$ 
\be\label{FQexp}
\cF=h(M)\ +\ f_{IJ}(M)\, Q^I Q^J+\ldots\ .
\ee
Inserted into (\ref{Kspecial2}) one finds \cite{jl}
\be
K=\hat{K}(M,\bar{M})\ +\ Z_{IJ}(M,\bar{M})\,
 Q^I\bar{Q}^{J}+\ldots
\ee
where 
\bea\label{KZdef}
\hat{K}&=&-\ln\Big( 2 (h+\bar{h})-(M^i-\bar{M}^i) (h_i-\bar{h}_i)\Big)\nonumber\\
Z_{IJ}&=& 4\, e^{\hat{K}}\R f_{IJ}(M)\ .
\eea

The gauge couplings of any 
non-Abelian factor $G_a$ also simplify since  gauge invariance of 
eq.~(\ref{Lgauge}) requires $f_{IJ}=\del_{IJ}\, f_a$ 
where here $I$ and $J$ label
the vector multiplets of the factor $G_a$.
Inserted into eqs.~(\ref{gdef}),(\ref{Ndef}) reveals that the (Wilsonian) gauge coupling of
a non-Abelian factor is a harmonic function of the moduli
\be\label{gtwotree}
g^{-2}_a=\R f_a(M)\ .
\ee
As for $N=1$ this only holds at the tree level.
At the loop level the effective (non-Abelian)
gauge couplings again cease to be 
harmonic and instead obey the $N=2$ analog of   
eq.~(\ref{ftg}) which is found to be \cite{jl}
\be\label{gtwo}
g_a^{-2} = \R f_a 
+\frac{b_a}{16\pi^2} \Big(\ln\frac{M_{Pl}^2}{p^2}+ \hat{K}(M,\hat{M})\Big).
\ee
One way to derive this relation is to
simply insert eqs.~(\ref{KZdef}), (\ref{gtwotree})
 into eq.~(\ref{ftg}).\footnote{As for
$N=1$ vacua this formula can also be viewed 
as an all-loop expression.
In $N=2$ the $\beta$-function is only corrected at
one-loop in perturbation theory but not beyond.}

The moduli of $N=2$ supergravity can be 
scalars in either vector  or hypermultiplets
and the total moduli space is a locally a direct 
product as in eq.~(\ref{modulispace}). 
However, from eq.~(\ref{gtwo}) we learn that
the effective gauge couplings only depend on the 
moduli $M^i$ in vector multiplets
but not on  the moduli in hypermultiplets. 
This is a consequence of the fact that the $N=2$ gauge
fields couple to charged hypermultiplets 
in a minimal gauge covariant way
but they do not have two derivative
couplings with any of the neutral hypermultiplets 
\cite{wlp} and hence the gauge couplings 
also cannot depend on the latter.
However,  the spectrum of charged hypermultiplets does
enter into eqs.~(\ref{gtwo}) in that they affect 
the $\beta$-function coefficients $b_a$.

\subsection{$N=2$ Heterotic Vacua}
So far we reviewed the effective 
$N=2$ supergravity without any reference to a particular 
string theory. The aim of this section is 
to determine the prepotential $\cF$ for 
$N=2$ heterotic vacua.\footnote{The presentation of this section
closely follows ref.~\cite{jl}; similar results were 
obtained in ref.~\cite{afg}. 
(See also refs.~\cite{cdadf,cdafvp}.) }

\subsubsection{$N=2$ Non-Renormalization Theorems}
As we already discussed,  for  heterotic vacua 
the dilaton is part of a  vector-tensor multiplet  
but we choose to discuss it in terms of its 
dual  vector multiplet. 
More precisely, the dilaton is the real part of the 
scalar component in the dual vector multiplet with
the axion being the imaginary part. 
From the fact that the dilaton organizes 
the string perturbation theory
together with product structure of the moduli 
space (\ref{modulispace}) one 
derives the following
non-renormalization theorem \cite{jl,strominger}: 
\begin{itemize}
\item[(i)]
The  moduli space of the hypermultiplets 
is determined at the string tree level and 
receives no further perturbative or non-perturbative
corrections, i.e.
\be
\cM_H=\cM_H^{(0)}\ .
\ee
\end{itemize}
A.~Strominger \cite{strominger}
stressed the validity of  this non-renormalization
theorem beyond string perturbation theory. 
This is a consequence of the
fact that this theorem
only depends on unbroken $N=2$ supersymmetry and the 
assignment of the dilaton multiplet. 
In particular it 
does not depend on the continuous 
PQ-symmetry
which in part was responsible 
for the non-renormalization
theorem of  $W$ and $f_a$ in
$N=1$ heterotic vacua.
However, it was precisely the breaking of the 
PQ-symmetry which allowed for a violation 
of the $N=1$ non-renormalization theorem
by non-perturbative effects.

On the other hand the PQ-symmetry 
(\ref{PQ}) can be used to derive a second
non-renormalization theorem in $N=2$.
The loop corrections of the prepotential $\cF$
are organized in 
an appropriate power series expansion in the dilaton.
But exactly as in the $N=1$ case the holomorphy of $\cF$
and the PQ-symmetry only allow a very limited number of terms. 
Repeating the analysis of section~2.2.~one finds:
\begin{itemize}
\item[(ii)] 
The prepotential $\cF$
only receives contributions at the string tree
level $\cF^{(0)}$, at one-loop $\cF^{(1)}$ 
and non-perturbatively $\cF^{(NP)}$, i.e.
\bea\label{Floopexp}
\cF&=&\cF^{(0)} (S,M)\ +\ \cF^{(1)}(M)\nonumber\\
& & +\ \cF^{(NP)}(e^{-8 \pi^2 S},M)\ .
\eea
\end{itemize}
$\cF^{(0)}$ is   of  order $S$, 
$\cF^{(1)}$ is dilaton independent and 
$\cF^{(NP)}$ is  only
constrained by the discrete PQ-symmetry (\ref{PQdis}).
This second theorem is the `cousin' of the 
$N=1$ non-renormalization theorem 
for  $W$ and $f_a$ but  in $N=2$ it is more 
powerful since $\cF$ determines both the Wilsonian
gauge coupling and the K\"ahler potential.

\subsubsection{String Tree Level}
The next step is 
to determine the tree level prepotential $\cF^{(0)}$
using additional information that 
we have at our disposal.
All (perturbative) $N=2$ heterotic vacua also satisfy 
\begin{itemize}
\item The tree level gauge coupling of any non-Abelian
factor $G_a$  is the set by the VEV of the dilaton.\footnote{
This only holds for the perturbative gauge group.
In the last lecture we will see that non-perturbative
effects can enlarge the gauge group but with a
different coupling to the dilaton \cite{witten}.}
That is, $g^{-2}= \R S$ or 
\be\label{ftree}
f^{(0)}\ =\ S.
\ee
\item  The dilaton dependence of the tree level K\"ahler potential 
is constrained by the PQ-symmetry 
and the fact that the dilaton arises in the universal
sector. Therefore  it cannot  mix with any other scalar field
at the tree level and one necessarily has
\be
K^{(0)}=-\ln(S+\bar{S})+\tilde{K}(M,\bar{M}, Q, \bar Q) .
\ee
\end{itemize}
Surprisingly, this separation of the dilaton piece together with the constraint 
(\ref{Kspecial2})   uniquely fixes $\cF^{(0)}$ 
to be \cite{fpr}:
\be\label{Ftreehet}
\cF^{(0)}=- S\Big( \eta_{ij}  M^i M^j-\del_{IJ}  Q^I Q^J\Big)\ ,
\ee
where $\eta_{ij}={\rm diag}(1,-1,\ldots,-1)$.
Inserted into  (\ref{Kspecial2}) 
the tree level K\"ahler potential is found to be
\bea
K^{(0)}&=&\! -\ln(S+\bar{S})\\
&-&\! \ln\Big(\eta_{ij}   \R M^i\R M^j
- \del_{IJ}  \R Q^I\R Q^J\Big). \nonumber
\eea
The metric derived from this K\"ahler potential
is the metric of the coset space 
\be
\cM_V^{(0)}=\frac{SU(1,1)}{U(1)}\times 
\frac{SO(2,n_V-1)}{SO(2)\times SO(n_V-1)}\ ,
\ee
where the first factor of the moduli space is 
spanned by the dilaton and the second factor
by the other vector multiplets. 
Let us stress once more that this results generically 
holds for  perturbative heterotic
$N=2$ vacua and is a consequence of the constraints implied by
supergravity and the special properties of the dilaton couplings.

\subsubsection{Perturbative Corrections}
The next step is to determine $\cF^{(1)}$, i.e.~the one-loop corrections 
of the prepotential.  
Inserting (\ref{FQexp}) into (\ref{Floopexp}) we see that both 
$h$ and $f_{IJ}$ have their own  loop expansion
\bea
h&=&h^{(0)}+\ h^{(1)}+\ h^{(NP)}\ ,\nonumber\\
f_{IJ}&=&f_{IJ}^{(0)}+\ f_{IJ}^{(1)} +\ f_{IJ}^{(NP)}\ ,
\eea
and thus one has to 
compute   $f^{(1)}$ and $h^{(1)}$.
However, for these two couplings  one cannot derive a result
as general as we just did for $\cF^{(0)}$. 
Instead we again only consider a 
particular subclass of heterotic $N=2$ vacua 
and apply the method developed in the previous section.
In order to use some of the earlier results 
we focus on those vacua  which have
two moduli $T$ and $U$ and a  perturbative 
quantum symmetry
$SL(2,{\bf Z})_T\times SL(2,{\bf Z})_U$  
acting on $T$ and $U$ as in eq.~(\ref{modtr}).
To be slightly more specific, let us consider
compactifications of the ten-dimensional heterotic string
on $K3\times T^2$ where
$T$ and $U$ are the two
toroidal moduli of $T^2$ and $K3$ is a 
four-dimensional Calabi--Yau manifold.\footnote{
There are vacua which have an 
$SL(2,{\bf Z})_T\times SL(2,{\bf Z})_U$ 
quantum symmetry but which cannot be
interpreted as geometrical compactifications.
We do not  get into these subtleties here.}

$T$ and $U$ are the scalar components of two Abelian 
 vector multiplets so that 
for this class of vacua  the gauge group is
\be
G=G'\times U(1)_T\times U(1)_U\times U(1)_S\times U(1)_{\gamma}\ .
\ee
$G'$ is a gauge factor which we do  not further
specify since it depends on the 
particular vacuum under consideration.\footnote{$G'$ 
also varies over the moduli space.
As we already remarked, in $N=2$ the vector multiplets
in the Cartan subalgebra of $G'$ are flat directions
and therefore away from the origin of moduli space
$G'$ is generically broken to some subgroup.
For example, compactification on $K_3\times T^2$
in which the spin connection is embedded in the gauge connection
of $K_3$ in the standard way,
has a  gauge group
$G=E_8\times E_7\times U(1)_T\times U(1)_U\times U(1)_S\times U(1)_{\gamma}$
at the origin of the vector multiplet moduli space 
which at a generic point is broken to 
$G=U(1)^{15}\times U(1)_T\times U(1)_U\times U(1)_S\times U(1)_{\gamma}$.
For this family of vacua there are also $65$ moduli in hypermultiplets.}
However, the rank of $G$ is again bounded 
by the central charge  $\bar c_{int}=22$.
For $N=2$ vacua there are two additional $U(1)$ 
gauge bosons in the universal sector
corresponding to the graviphoton and the 
superpartner of the dilaton.
Thus we have
\be\label{rankhet}
{\rm rank}(G)\ \le\ 22+2\ .
\ee

$h^{(1)}(T,U)$ and $f^{(1)}(T,U)$ can be determined by an appropriate
string loop computation.
Here we follow instead the same method 
developed for  $N=1$ vacua 
and use the quantum symmetries and 
singularity structure to
determine $h^{(1)}(T,U)$ and $f^{(1)}(T,U)$.
However, there is slight complication now since
$h^{(1)}(T,U)$ does have singularities inside
the fundamental domain. As we already stated
earlier there are additional gauge neutral
massless modes on the subspace $T=U$ and in 
$N=2$ vacua they belong to Abelian vector multiplets.
Such states induce a logarithmic singularity
in the $U(1)$ gauge couplings and by supersymmetry
also  render the associated $\theta$-angles 
ambiguous. From eqs.~(\ref{Lgauge})--(\ref{FQexp})
we learn that $h$ is directly related to the
$U(1)$ gauge couplings while $f_a$ encodes
the gauge couplings of the non-Abelian factors $G_a$. 
As explained in the previous section the $f_a$ 
are non-singular inside the fundamental domain.

In order to determine the transformation properties
of $h^{(1)}(T,U)$ under the modular group one
can use a general formalism developed in ref.~\cite{jl}.
Here we just state the result that  $h^{(1)}$
has to be a modular form of weight $-2$ if it 
were nowhere  singular.
In the presence of singularities one has to allow
for integer ambiguities of the $\theta$-angles which results in
\be\label{htrans}
h^{(1)}(T,U)\to \frac{h^{(1)}(T,U)+\Xi(T,U)}{(i c T+d)^2}
\ee
for an $SL(2,{\bf Z})_T$ transformation.
(For $SL(2,{\bf Z})_U$  $T$ and $U$
are interchanged in eq.~(\ref{htrans}).)
$\Xi$ is an arbitrary quadratic polynomial in the 
variables $(1,i T, i U, TU)$  and parameterizes
the most general allowed ambiguities in the
$\theta$-angles; $\Xi$ obeys
\be\label{xiprop}
\p_T^3\Xi=\p_U^3\Xi=0\ .
\ee
As we said, for $\Xi=0$ 
$h^{(1)}$ is a modular  form
of weight $(-2,-2)$ with respect to 
$SL(2,{\bf Z})_T\times SL(2,{\bf Z})_U$
but for non-zero  $\Xi$ it has no good modular
properties. However, from 
eqs.~(\ref{xiprop}), (\ref{modder}) we learn that instead 
$\p_T^3 h^{(1)}$ is a single valued modular form of 
weight $(4,-2)$  and similarly 
$\p_U^3 h^{(1)}$ has weight $(-2,4)$.

The singularities in the $T,U$ moduli space arise 
along the critical line $T=U$ (mod $SL(2,{\bf Z})$)
where two additional massless gauge fields appear and the $U(1)_T\times U(1)_U$ is
enhanced to $SU(2)\times U(1)$.
Further enhancement appears at $T=U=1$, 
which is the intersection of the two critical 
lines $T=U$ and $T=1/U$. In this case one has $4$ 
extra gauge bosons and an
enhanced gauge group $SU(2)\times SU(2)$. 
The intersection at the critical point $T=U=\rho=e^{2\pi i/12}$ gives rise to $6$ massless gauge bosons 
corresponding to the gauge group $SU(3)$.
Altogether we have:
\begin{center}
\bt{ll}
$T=U$:& $\!\!\! U(1)_T\times U(1)_U \to SU(2)\times U(1)$\\
$T=U=1$:& $\!\!\! U(1)_T\times U(1)_U \to SU(2)\times SU(2)$\\
$T=U=\rho$:& $\!\!\! U(1)_T\times U(1)_U \to\ SU(3)$\\
\et
\end{center}

The singularity of the prepotential near $T=U$ can be determined
by purely field theoretic considerations \cite{seib}.
For the case at hand one finds \cite{seib,sw,jl}
\bea\label{hsing}
h^{(1)} (T\sim U)&=&\frac{1}{16\pi^2}\,(T-U)^2\,
\ln(T-U)^2\nonumber\\
& &\ +\ {\rm regular\ terms}
\eea
where the coefficient $\frac{1}{16\pi^2}$ is set by the 
$SU(2)$ $\beta$-function. 
(The general derivation of eq.~(\ref{hsing}) is reviewed
in detail in the lectures by W.~Lerche; see also \cite{jl}.)
From eq.~(\ref{hsing}) we learn that 
$\p_T^3 h^{(1)}$ and $\p_U^3 h^{(1)}$ have a simple pole at $T=U$.

As we already discussed
in section 2, in the decompactification limit ($T,U\to \infty$)
the gauge coupling cannot grow faster than a single
power of $T$ or $U$ and the exact
 same argument also applies for $h^{(1)}$. 
Since the moduli dependence of the gauge coupling
is related to the second derivative of $h$ we learn that 
$\p_T\p_U h^{(1)}$, $\p^2_T h^{(1)}$
and $\p^2_U h^{(1)}$ cannot grow faster than 
$T$ or $U$ in the decompactification limit.
Hence
\be
\p_T^3 h^{(1)}\to {\rm const.}\ , \quad \p_U^3 h^{(1)}\to {\rm const.}\ 
\ee
for either $T\to \infty$ or $U\to \infty$.
The properties of  $h^{(1)}$ which we have
assembled so far can be
combined in the Ansatz 
\be\label{ansatz}
\p_T^3 h^{(1)}=\frac{X_{-2}(U)\,  Y_4(T)}{j(i T)-j(i U)}\ ,
\ee
where $X_{-2}(U)$ and $Y_4(T)$ are modular forms of weight 
$(0,-2)$ and $(4,0)$, 
respectively.  $X_{-2}(U)$ and $Y_4(T)$
cannot have any pole inside the modular domain 
while for large $T,U$ they 
have to obey
\bea
U\to \infty : & \frac{X_{-2}(U)}{j(i U)}&\to {\rm const.}\ ,\nonumber\\
T\to \infty : & \frac{Y_4(T)}{j(i T)}&\to {\rm const.}\ .
\eea
From the theory of modular forms
(see appendix~A)
one infers that these properties uniquely determine 
$X_{-2}=E_4 E_6/\eta^{24}$ and $Y_4=E_4$;
inserted into (\ref{ansatz}) yields
\be
\p_T^3 h^{(1)}\ =\ \frac{1}{2\pi}
\frac{E_4(i T)\, E_4 (i U)\, E_6 (i U)}{[j(i T)-j(i U)]\, \eta^{24}(i U)}\ ,
\ee
where the coefficient is determined by eq.~(\ref{hsing}) or rather the 
$SU(2)$ $\beta$-function.
The same analysis repeated for $\p_U^3 h^{(1)}$ reveals
\be
\p_U^3 h^{(1)}=\frac{-1}{2\pi}
\frac{E_4(i U)\, E_4 (i T)\, E_6 (i T)}{[j(i T)-j(i U)]\, \eta^{24}(i T)}\ .
\ee

The analysis just performed only determines the third derivatives
of $h^{(1)}$ because these are modular forms.
$h^{(1)}$ itself has been calculated in 
ref.~\cite{harmo} by explicitly calculating
the appropriate string loop diagram.\footnote{See also 
refs.~\cite{k,KKPR,DMVV,HM,ccl,harmo2}.}
An intriguing relation with hyperbolic Ka\v c--Moody algebras and 
Borcherds denominator formula was found.
Unfortunately, a review of these exciting developments
is beyond the scope of these lectures.

In addition to the transformation law of $h$ (eq.~(\ref{htrans}))
the general formalism of ref.~\cite{jl} also reveals 
that the $N=2$ dilaton is no longer invariant at the quantum level.
Instead, under an  $SL(2,{\bf Z})_T$ transformation one finds
\be\label{Strans}
S\to S+\hf\p_T\p_U \Xi- 
\frac{ic\p_U (h^{(1)}+\Xi)}{2(i c T+d)} 
 + {\rm const.}
\ee
This somewhat surprising result can be understood from the fact that
in perturbative string theory the relation between the dilaton and the
vector-tensor multiplet is fixed. 
However, the duality relation between the vector-tensor multiplet
and its dual vector multiplet containing
$S$ is not fixed but suffers from perturbative corrections 
in both field theory and string theory. 
Nevertheless,
it is possible to define an invariant dilaton by
\be\label{Sinv}
S^{\rm inv}\equiv S-\hf\p_T\p_U h^{(1)}
-\frac{1}{8\pi^2}\ln[j(iT) - j(iU)]\ .
\ee
The last term is added such that $S^{\rm inv}$ is finite
so that altogether
$S^{\rm inv}$ is modular invariant and finite.
However,  $S^{\rm inv}$
it is no longer an $N=2$ special coordinate.\footnote{
In $N=1$ it is always possible to keep the dilaton
superfield chiral and modular invariant by an
appropriate holomorphic field redefinition.}

We are now ready to determine the one-loop correction $f^{(1)}$.
As before we demand that the effective gauge couplings (\ref{gtwo})
remain invariant under $SL(2,{\bf Z})_T\times SL(2,{\bf Z})_U$. 
Using (\ref{gtwo}),(\ref{f}),(\ref{Ktrans}) one finds 
that this requires the transformation property
\be\label{ftwotrans}
f_a(S,T,U) \to f_a(S,T,U)-\frac{b_a}{8\pi^2}\ln(i c T+d)\ ,
\ee
where $f_a$ is the entire function 
and not only the one-loop contribution.
Using eqs.~(\ref{Strans})-(\ref{ftwotrans})
together with the  tree level contribution  
(\ref{ftree}) one finds (again up to a constant)
\be
f_a(S,T,U) =  S^{\rm inv}
-\frac{b_a}{8\pi^2}\ln[\eta^2(i T)\eta^2(i U)]
\ee
or equivalently
\be
f^{(1)}_a=\frac{-b_a}{8\pi^2}\ln[\eta^2(iT)\eta^2(iU)]
   +  (S^{\rm inv}-S).
\ee
 
To summarize, as for the $N=1$ factorizable orbifolds
we managed to use the perturbative quantum symmetries 
and the singularity structure to determine the one-loop
correction of the gauge couplings.
For the $U(1)$ factors we did not completely
determine $h^{(1)}$ but only its  third derivatives.
$h^{(1)}$ itself can be computed using the formalism
developed in ref.~\cite{harmo}.
Further generalization to different classes of 
$N=2$ string vacua can be found for example in
refs.~\cite{afg,ms,AP,harmo,k,HM,ccl}.

Up to now we mostly concentrated on the supersymmetric gauge couplings 
because of their special analytic properties.
In addition,
there is a class of higher derivative curvature terms
whose couplings $g_n$ are also determined
by holomorphic functions $\cF_n(S,M^i)$ of the dilaton 
and the moduli \cite{bcov,agntop,dewit}.\footnote{The
prepotential $\cF$ as well as these higher derivative couplings 
$\cF_n$ arise from chiral integrals in 
$N=2$ superspace.}
(The properties of these couplings are the subject of 
the lectures by K.S.~Narain.)
More specifically, terms of the type 
\be\label{gndef}
\cL \sim g_n^{-2} R^2 G^{2n-2} + \ldots\ ,
\ee 
where $R$ is the Riemann tensor and $G$ the field strength 
of the graviphoton
are governed by couplings $g_n$ that are almost harmonic
\be\label{Fndef}
g_n^{-2}=\R \cF_n(S,M^i)+\cA_n\ .
\ee
At the tree level $A_n=0$ holds and 
thus $g_n^{-2}$ is a harmonic function. 
However exactly as for the gauge coupling 
the $g_n^{-2}$ cease to be harmonic as soon as
quantum corrections are included, or in words
a holomorphic anomaly $A_n\neq 0$ is induced \cite{bcov}.
The particular form of the anomaly is not of
immediate concern here but can be found for example
in refs.~\cite{bcov,vjs,agnt}  and K.S.~Narain's lectures.
The same argument we gave earlier for $f_a$ and
the prepotential $\cF$ implies that also the couplings 
$\cF_n(S,M^i)$ can be expanded in powers of the dilaton
and they have to respect the PQ-symmetry.
Thus, one has analogously 
\bea
\cF_n&=&\cF_n^{(0)}(S,M^i)+\cF_n^{(1)}(M^i)\nonumber\\
& &\  +\ \cF_n^{(NP)}(e^{-8 \pi^2 S},M^i)\ ,
\eea
where $\cF_n^{(0)}$ is the tree level term, 
$\cF_n^{(1)}$ the one-loop correction and 
$\cF_n^{(NP)}$ the non-perturbative contributions. 
Furthermore, the continuous PQ-symmetry (\ref{SPQ}) 
only allows a dilaton dependence for $n=1$ 
and one finds
\be\label{Fhigher}
\cF_1^{(0)}=24\, S,\quad  \cF_{n>1}^{(0)}={\rm const.}\ ,
\ee
where for $\cF_1^{(0)}$ a convenient normalization 
has been chosen.
For specific string vacua, some of the $\cF_n$ have
been computed in 
refs.~\cite{vjs,agnt,curio,cclmr,dwclm,HM}.

\subsection{Left-Right Symmetric Type II Vacua}

So far we focussed on heterotic $N=2$ vacua and in particular
on the moduli space of their vector multiplets.
Now we shift our attention to type II 
vacua but with the additional input that there 
is a symmetry between the left
and right moving CFT; this assumption
considerably simplifies the analysis.\footnote{More
general type II vacua are discussed for example in 
refs.~\cite{DKV,VW}.}
In $d=4$ such vacua have a universal sector with 
$c_{\rm st}=\bar{c}_{\rm st}=6$ that  contains
the gravitational multiplet and
the dilaton multiplet
which we already discussed in section~3.1.~and table~8.
The internal sector has 
$c_{\rm int}=\bar{c}_{\rm int}=9$ 
and if it is left-right symmetric
it also has $(2,2)$ global worldsheet supersymmetry. 
This corresponds to a  compactification of the ten-dimensional
type II string on a Calabi--Yau threefold.
(Calabi--Yau manifolds are reviewed in the lectures
by R.~Plesser; see also 
appendix~B and for example refs.~\cite{GSW,theisen1}.)

The massless spectrum of a type II vacuum
compactified on a Calabi--Yau threefold $Y$ 
is characterized by the Hodge numbers 
$h_{1,1}$ and $h_{1,2}$ (appendix~B).
For type IIA one finds \cite{s,cfg}
$h_{1,1}+h_{1,2}$
complex massless scalar fields in the $NS-NS$
sector and 
$h_{1,1}$ Abelian vectors together 
with $h_{1,2}$
complex scalars in the $R-R$ sector.
These states (together with their fermionic
partners) combine into 
$h_{1,1}$ vector multiplets
and $h_{1,2}$ hypermultiplets.
The total number of multiplets therefore is 
$n_V=h_{1,1},\ n_H=h_{2,1}+1$ 
where the extra hypermultiplet counts 
the dilaton multiplet of the universal sector.
For type IIB vacua 
one also has $h_{1,1}+h_{1,2}$
complex massless scalar fields in the $NS-NS$
sector but now 
$h_{1,2}$ Abelian vectors together 
with $h_{1,1}$
complex scalars in the $R-R$ sector \cite{s,cfg}.
Hence,
$n_V=h_{2,1}$ and  $n_H=h_{1,1}+1$ holds
for the type IIB theory.
The gauge group is always Abelian and given
by $(h_{1,1}+1)$ $U(1)$ factors in type IIA
and $(h_{1,2}+1)$ $U(1)$ factors in type IIB
(the extra $U(1)$ is the graviphoton in the
universal sector).
We summarize the spectrum of type II vacua in
table~9.

\begin{table}
\begin{center}
\begin{tabular}{|l|l|l|}\hline
{}&$IIA$&$IIB$ \\\hline
$n_H$&$h_{1,2}+1$&$h_{1,1}+1 $\\
$n_V$&$h_{1,1}$&$h_{1,2}$\\
$G$&$U(1)^{h_{1,1}+1}$&$U(1)^{h_{1,2}+1}$\\\hline
\end{tabular}
\end{center}
\caption{Massless spectrum in type II vacua.}
\end{table}

As we see the role of $h_{1,1}$ and $h_{1,2}$
is exactly interchanged between type IIA 
and type IIB. Therefore,
compactification of type IIA on 
a Calabi--Yau threefold $Y$ is equivalent to 
compactification of type IIB on the mirror
Calabi--Yau $\tilde Y$ (see appendix~B).
This is an example of a perturbative 
equivalence of two entire classes of string vacua.
The recent developments about string dualities
show that many more of such equivalences 
do exist but they often only hold when the
non-perturbative corrections are also taken into account. (See lectures by S.~Kachru, R.~Plesser
and J.~Schwarz.) We briefly return to this aspect in 
the next lecture.

We already discussed 
the geometry of the moduli space
of  $N=2$ supergravity in section~3.1.
For type II vacua there is a special feature
at the string tree level.
The $h_{1,1}$ and $h_{1,2}$ complex scalars
by themselves each form a special K\"ahler
manifold. This is an immediate consequence 
of table~9 in that they can both can be members 
of vector multiplets and thus their geometry
must be special K\"ahler. However, when they 
sit in hypermultiplets they pair up with
additional scalars from the $R-R$ sector
and this renders the combined geometry
quaternionic. Thus it cannot be an
arbitrary quaternionic geometry but
must originate from a 
special K\"ahler geometry.
This fact has been termed the $c$-map
in ref.~\cite{cfg} and it implies that
also this `special' quaternionic geometry 
is characterized by a holomorphic
prepotential.\footnote{The special
quaternionic geometry is also
called `dual  quaternionic'.}

However, the special quaternionic geometry 
only appears at the string tree level.
This is due to the fact that
in type II vacua the dilaton resides in a 
hypermultiplet
and thus there is an analogous `type II' version
of the non-renormalization theorem (i) 
of  section~3.2.1. It states
that the moduli space of the vector multiplets
is exact and not corrected either perturbatively
or non-perturbatively while $\cM_H$ does
receive quantum corrections.
In other words, for type II vacua one has\footnote{
As in the heterotic case, this non-renormalization theorem also holds non-perturbatively.}
\be
\cM_V = \cM_V^{(0)}\ .
\ee

The quaternionic geometry 
of both type II and heterotic vacua has not been
studied in much detail\footnote{See, however,
refs.~\cite{s,cfg,cecotti,fs,wpquat,ss,adfft,daff,abc,abcm}.}
and most investigations focussed on the special
K\"ahler geometry of the vector multiplets.
For Calabi--Yau threefolds $Y$  the prepotential $\cF$
enjoys some additional properties.
Of particular interest is the expansion around
geometrically large Calabi-Yau manifolds which is 
also referred to as the large radius 
or rather large volume limit.
In this limit the position 
in $\cM_V$ is given by the complexified K\"ahler form $B+i J$ which can be decomposed  into a basis 
$e_{\alpha}\in H^{1,1}(Y,{\bf Z})$ 
according to 
\be
B+i J=\sum_{\alpha=1}^{h_{1,1}} (B+i J)_{\alpha} e_{\alpha}=t_{\alpha} e_{\alpha}\ ,
\ee
where $B_{\alpha}$ are the moduli corresponding to the antisymmetric tensor and $J_{\alpha}$ the moduli which
are associated to the deformation of the metric;  
$t_{\alpha}$ are the corresponding complex
parameters ($\alpha=1,\ldots, h_{1,1}$)
which are also $N=2$ special coordinates.
In the large radius limit the moduli $B_{\alpha}$ 
are periodic variables and thus enjoy a discrete 
`PQ-like' symmetry. More precisely, one finds
that the low energy effective theory is invariant 
under 
$B_{\alpha}\to B_{\alpha}+1$ or equivalently
\be\label{PQt}
t_{\alpha}\to t_{\alpha}+1\  .
\ee
In the large volume limit 
the  prepotential obeys
\bea\label{FCYexp}
\cF &=&-\frac{i}{6}\, d_{\alpha\beta\gamma}\,
 t^{\alpha}t^{\beta}t^{\gamma} \nonumber\\
& &+\ \ {\rm worldsheet\;\; instantons}\ ,
\eea
where $d_{\alpha\beta\gamma}$ are the 
classical intersection numbers of the 
$(1,1)$ forms defined by
\be
d_{\alpha\beta\gamma}:=\int_{Y} e_{\alpha}\wedge e_{\beta}\wedge e_{\gamma}\ .
\ee
Using mirror symmetry \cite{dixon,cls,GP}
the entire prepotential $\cF$ can be computed
for string vacua with small number of vector multiplets
(see for example \cite{CDGP,morrison,font,KT,hkty,BK,cofkm,cfkm,hkty2,BKK}).
In the limit $t_{\alpha}\to\infty$ the 
contributions from worldsheet instantons vanish.

The higher derivative couplings $\cF_n$ which 
were defined in eqs.~(\ref{gndef}),(\ref{Fndef})
and which similarly only depend on the vector multiplets,
also obey the type II non-renormalization theorem.
For these couplings one finds  the additional 
curiosity that they are proportional to the 
genus $n$ 
topological partition functions of a
twisted Calabi--Yau  $\sigma$--model \cite{bcov}.
That is, they receive a contribution only at 
a fixed string loop order
and satisfy certain recursion relations expressing 
the holomorphic anomaly of a
genus $n$ partition function to the
partition function  of lower 
genus \cite{bcov}.
The properties of these couplings both in type II
and the heterotic string are reviewed in detail 
in K.S.~Narain's lectures.
In the large radius limit they obey
\bea\label{FCYhigher}
\cF_1&=&-\frac{i}{4\pi}\sum_{\alpha} t^{\alpha}\int_{Y} e_{\alpha}\wedge c_2(Y)\nonumber\\
& & +\ {\rm worldsheet\;\;instantons}\\
\cF_{n>1}&=&{\rm const.}\ +\
{\rm worldsheet\;\;instantons}\nonumber
\eea
where $c_2$ is the second Chern class of the 
Calabi--Yau manifold.

%

\section{Heterotic-Type II Duality}
So far we exclusively focussed on perturbative
properties of string vacua although we noticed 
a number of non-renormalization theorems 
which also hold non-perturbatively.
During the past two years it has become clear that
there are much more intricate relations between
(classes of) different string vacua than 
had hitherto been imagined.
In particular, it is believed that string vacua
which look rather different in string perturbation
theory can yet be equivalent when all 
(perturbative and non-perturbative) quantum 
corrections are taken into account.
By now there is whole web of relations among
perturbatively distinct string vacua.
These relations strongly depend on the amount
of space-time supersymmetry and the number
of space-time dimensions.
This web of interrelations has been 
discussed in detail in the lectures of J.~Schwarz.
Here we focus on one particular correspondence
namely the heterotic -- type II duality in $d=4$
with $N=2$ supersymmetry. This duality 
is also the subject of the lectures of S.~Kachru
and in this closing section we only briefly
outline how the perturbative properties
found earlier have been used to support 
the conjecture of this  particular duality.

The conjecture states 
that $N=2$ heterotic vacua are  quantum 
equivalent to $N=2$ type IIA vacua
and vice versa \cite{kv,fhsv}.\footnote{Since there
is already the perturbative relation that
type IIA vacua compactified on a Calabi--Yau
manifold $Y$ are equivalent to  type IIB vacua
compactified on the mirror $\tilde Y$ one really
has a triality heterotic $\sim$ type IIA 
$\sim$ type IIB where the first equivalence 
only holds non-perturbatively while the second
is a perturbative equivalence.}
On face value this conjecture seems unlikely to hold.
First of all, for heterotic vacua the rank
of the gauge group is bounded by the central charge
to be less than 24 (eq.~(\ref{rankhet}))
while in type II vacua the rank can certainly be
much larger since both Hodge numbers easily exceed
22 ({\sl c.f.}~table 9).
Furthermore, we saw that the heterotic vacua have large 
non-Abelian gauge groups at special points in their
moduli space while type II A vacua only have an
Abelian gauge group.
However, the analysis of Seiberg and Witten \cite{sw}
taught us that asymptotically free
non-Abelian gauge groups
generically do not survive non-perturbatively but instead 
are broken to their Abelian 
subgroups.\footnote{
Conversely one has to show that in a particular limit
corresponding to the heterotic weak coupling limit
a type II vacuum can have a non-Abelian enhancement
of its gauge group \cite{aspinwall,LAwitten}.}

It is not completely clear yet that the 
heterotic--type II duality holds
for the whole space of heterotic and type II vacua
or only on a (well defined) subspace.
So far specific examples of string vacua
or classes of string vacua have been proposed 
to be non-perturbatively equivalent.
For such vacua -- which are also called a `dual pair' --
the low energy effective theories 
have to be identical when all quantum corrections are
taken into account.
In particular their moduli spaces have to  coincide,
i.e.
\bea
\cM_V^{het}&=&\cM_V^{IIA}\ ,\nonumber\\
\cM_H^{het}&=&\cM_H^{IIA}\ .
\eea
Since the prepotential depends on the vector multiplets in both theories the equality of 
the moduli spaces 
amounts to the equality of the prepotential 
\be
\cF_{het}=\cF_{IIA}\ .
\ee
This should not only be true for the prepotential
but also hold for the higher derivative couplings
$\cF_n$. (See  K.S. Narain's lectures.)

The proof of this conjecture is problematic
since the prepotential on the heterotic side 
$\cF_{het}$ is only known perturbatively,
that is in a weak coupling expansion. 
For a heterotic vacuum weak coupling
corresponds to large $S$ and hence there 
has to be a type II modulus in a
vector multiplet which is identified
with the heterotic dilaton.
It is immediately clear that this type II modulus
cannot be the type II dilaton which always
comes in a hypermultiplet.
Instead it has to be one of the $h_{1,1}$
K\"ahler deformations of the Calabi--Yau threefold
in the large radius limit.\footnote{This 
immediately tells us that the appearance 
of the vector-tensor multiplet must be an
artifact of heterotic perturbation theory.
Similarly, the type II tensor multiplet 
is an artifact of type II perturbation theory.}
Thus one is interested in identifying 
this dual type II partner $t^s$
of the heterotic dilaton $S$.
From the discrete PQ-symmetries (\ref{PQdis}),(\ref{PQt}) 
one immediately infers
that the relation must be 
\be
t^s\equiv 4 \pi i S\ .
\ee

Once $t^s$ has been identified one can expand
the type IIA prepotential $\cF_{IIA}$ in a 
$t^s$ perturbation expansion around large $t^s$,
i.e.
\be
\cF_{II}=\cF_{II}(t^s,t^i)+\cF_{II}(t^i)
+\cF_{II}(e^{-2\pi i t^s},t^i)
\ee
where we use the notation $t^\alpha = (t^s,t^i)$.
This expansion can be compared to the 
perturbative expansion of the heterotic prepotential.
In particular one has to find
\be\label{dualcheck}
\cF_{II}(t^s,t^i)+\cF_{II}(t^i)
= \cF_{\rm het}^{(0)}(S,\phi^I)
 +\cF_{\rm het}^{(1)}(\phi^I)
\ee
(up to an overall normalization which 
is convention and has to be adjusted appropriately).
Let us reiterate that the left hand side of this 
equation is determined at the tree level 
whereas the right hand side sums 
perturbative contributions at the tree level and at
 one-loop.

Eq.~(\ref{dualcheck}) has been verified for
number of explicit string vacua.
Typically, these are vacua with a small number of
vector multiplets or low $h_{1,1}$
where the perturbative prepotential
is known on both sides.
These examples
can be found in S.~Kachru's
lectures or in 
refs.~\cite{vjs,klm,cclm,afiq,DM,cf,bkkm,lsty,k,ccl}.

Eq.~(\ref{dualcheck}) checks the spectrum
of the two vacua and their quantum symmetries.
However, it turns out that this does not uniquely identify a dual pair.
Instead, there can be an entire 
class of vacua where each member
satisfies eq.~(\ref{dualcheck}) but nevertheless
their 
non-perturbative prepotentials
$\cF^{(NP)}$ are different
\cite{MV,AG,bkkm,lsty}.
For  these cases additional, non-perturbative
information is necessary to uniquely identify
a dual pair. Unfortunately we cannot go into
any further detail about such examples.

It has also been shown that on the heterotic
side one discovers the Seiberg--Witten theory
in the field theory limit $\Mstring\to \infty$
\cite{kv,klm2,kklmv,clm2,AP}.
(This is reviewed in the lectures
by W.~Lerche.)
Finally, the matching of the higher derivative
gravitational couplings
can be found in refs.~\cite{vjs,agnt,curio,cclmr,dwclm} and K.S.~Narain's lectures.

Apart from the specific checks we just mentioned it 
is of interest to determine some more generic properties
of the heterotic--type II duality.
That is, one would like to study in general
the relation between a dual pair (or a class of dual
pairs) as well as the space of string vacua for which this 
conjectured duality holds.
From all our previous discussion it is clear that
generic properties should  involve the
heterotic dilaton since it couples
universally for all heterotic vacua.
Indeed, from 
eqs.~(\ref{Ftreehet}),(\ref{FCYexp}),(\ref{dualcheck}) 
one infers that the 
Calabi--Yau intersection numbers of a dual
type IIA vacuum have to obey
\be\label{cond1}
d_{sss}=0\ , \qquad d_{ssi}=0\quad \forall i\ ,
\ee
and 
\be\label{cond2}
{\rm sign}(d_{sij})=(+,-,\ldots,-)={\rm sign}(\eta_{ij})\ .
\ee
In addition, eqs.~(\ref{Fhigher}), (\ref{FCYhigher}) imply
\be\label{cond3}
\int e_S\wedge c_2(Y)=24\ .
\ee
These conditions are not unknown in the 
mathematical literature. They are  the
statement that the Calabi--Yau manifold
has to be a $K3$-fibration \cite{klm,al}.
That is, the Calabi--Yau manifold is fibred
over a ${\bf P^1}$ base with fibres that are $K3$
manifolds. The size of the ${\bf P^1}$ is parameterized
by the modulus $t^s$ which is the type II dual of the
heterotic dilaton.\footnote{
For a more detailed discussion about
$K3$-fibrations see for example \cite{klm,VW,al,HS}.}
Over a finite number of points on the base,
the fibre can degenerate to something other
than $K3$ and such fibres are called singular.
The other K\"ahler moduli $t^i$ are either moduli 
of the $K3$ fibre or of the singular fibres.
In general one finds
\be\label{genericd}
{\rm sign}(d_{sij})=(+,-,\ldots,-,0,\ldots,0)\ ,
\ee
where the non-vanishing entries correspond to moduli
from generic $K3$ fibres while
the zeros arise from singular fibres.
Since a $K3$ has at most 20 moduli
the non-vanishing entries have to be less than
$20$.\footnote{There is a possible subtlety
here since this counts only geometrical $K3$ moduli.
However, it is conceivable that quantum effects
raise this number up to 22 \cite{AM}.}
Comparing 
eqs.~(\ref{cond2}) and  (\ref{genericd})
one concludes that type II Calabi-Yau compactification
in the large radius limit can be the 
dual of perturbative heterotic vacua
if they are $K3$-fibration with all moduli
corresponding to generic fibres.
This class of type II vacua is automatically 
consistent with the heterotic bound
on the  rank of the gauge group (\ref{rankhet}).

Of course, one immediately asks the question
what can be the role of all the other type II
vacua. 
From eqs.~(\ref{Ftreehet}) and (\ref{genericd})
we learn that the 
$(1,1)$ moduli of singular fibres have no
counterpart in perturbative heterotic vacua.
If there were heterotic moduli with such couplings
they would  not couple properly 
to the (heterotic) dilaton and furthermore
violate the bound (\ref{rankhet}).
However, Witten observed that heterotic string vacua in
six space-time dimensions
have 
singularities when gauge instantons shrink to 
zero size \cite{witten}. 
He argued that these singularities are caused
by a non-perturbative enhancement of the
gauge group which 
opens up at the point of the zero-size instanton.
He also showed that these non-perturbative
gauge fields do not share the canonical
coupling to the dilaton.
In fact, upon compactification to $d=4$
one can show that the scalars of these 
non-perturbative vector
multiplets 
couple precisely like type II moduli corresponding
to singular fibres \cite{AG,lsty}.

This leads to the following general picture 
of the heterotic--type II duality.
\begin{figure}[htp]
\input{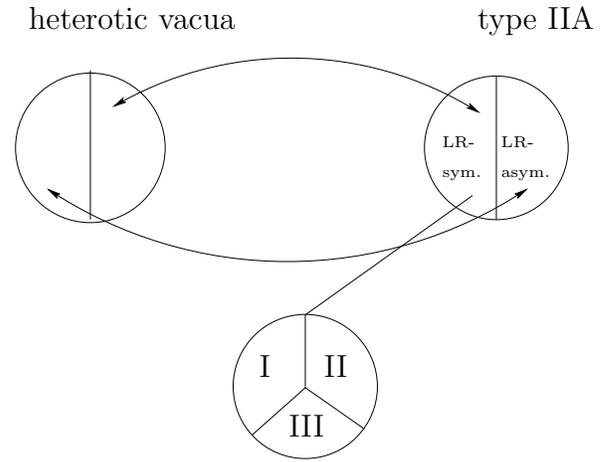}
\caption{Heterotic -- type II duality.}
\end{figure}
The space of left-right symmetric vacua
can be partioned into three different regions
(fig.~6).
\begin{itemize}
\item[I] $K3$--fibration with moduli only 
from generic fibres. 
All such vacua should be non-perturbatively
equivalent to 
perturbative heterotic vacua. 

\item[II] $K3$--fibration with moduli also from 
singular fibres. 
For such vacua a dual type II candidate for the
heterotic dilaton does exist 
but some of the other moduli do not couple 
perturbatively; they have to arise from 
vector multiplets which cannot be seen in
heterotic perturbation theory.

\item[III] Calabi--Yau manifolds which are
not $K3$--fibration. Such vacua have  no dual
candidate for the heterotic dilaton and thus they cannot 
be the dual of a weakly coupled heterotic vacuum.
However, they could well be the dual of 
heterotic vacua which have no weak coupling limit
with a dilaton that is frozen at strong coupling.
In that sense this might be the most interesting class of type II vacua.
\end{itemize}

From refs.~\cite{strominger,GMS}
we learned that
transitions among string vacua can occur once
non-perturbative states are taken into
the effective action near singularities of the moduli
space.
In the same spirit it is particularly
interesting to study transitions 
between vacua that cross any of the boundaries
in fig.~6.
Examples of such transitions have already been observed 
in 
refs.~\cite{KMP,klmvw} and are reviewed in the 
lectures of R.~Plesser.

\vskip .5cm
{\bf Acknowledgement}

J.L.~would like to thank his collaborateurs
P. Aspinwall, B. de Wit, L. Dixon, V. Kaplunovsky, D. L\"ust,
J. Sonnenschein, S. Theisen and S. Yankielowicz
for very fruitful and enjoyable collaboration on the topics
presented in these lectures.

J.L.~is supported in part by a Heisenberg fellowship 
of the DFG.
K.F.~is also supported by the DFG.

\begin{appendix}

\section{The modular group $\sltwo$}

The modular group $\sltwo$ \cite{gpr,apostol,serre,schoeneberg}  
enters string theory in various places.
First of all,  it is the group of reparametrizations of a genus one 
worldsheet which are not continuously connected to the identity.
In the definition of the 
physical one-loop scattering amplitudes this additional gauge freedom
has to be taken care of.
Topologically a one-loop string amplitude is a torus and the modular
group acts on the complex structure  of this torus. 
However, tori in space-time also enter in 
string compactifications as for example in the construction of orbifold 
vacua. As a consequence the modular group also appears as a (quantum)
symmetry of specific space-time effective theories.

The modular group is defined by the following transformation
on the complex modulus $T$
\be\label{modtr}
T \to {aT -ib\over icT + d}\ , \qquad ad - bc = 1\ ,
\ee
where $a,b,c,d \in {\bf Z}$.\foot{
It is common to choose a different convention for $T$
where real and imaginary part are exchanged. More precisely,
for $\tau = i T$ one has $\tau \to {a\tau +b\over c\tau + d}$.}
It has two generators $\cS,\cT$ which act as
\bea
\cT:T&\to& T+i\nonumber\\
\cS:T&\to& - 1/T
\eea

The transformation (\ref{modtr})
maps the $\R T > 0$ region in a rather complicated
way onto itself. However, one can define a fundamental region by the
requirement that every point of the $\R T > 0$ complex plane
is mapped into this region in a unique way.
One conventionally chooses:
\be 
 \Gamma=\{-\hf \le \I T \le \hf, \R T >0, |T|^2 > 1\}
\ee
as the fundamental domain (figure~7). 
No two distinct points in $\Gamma$ are equal under a modular transformation.
There are two fixed points 
of the map (\ref{modtr}) on this fundamental
domain, namely $T=1$ and $T=\rho\equiv e^{i\pi\over 6}$.

\begin{figure}[htp]
\input{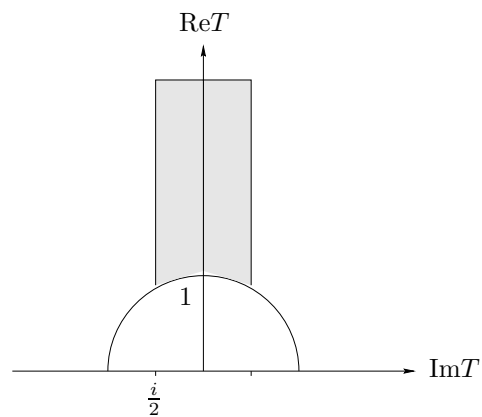}
\caption{Fundamental region $\Gamma$ of the modular group.}
\end{figure}

A modular form $F_r(T)$ of weight $r$ is defined to be holomorphic 
and to obey the transformation law
\be\label{modform}
F_r(T) \to (icT+d)^r F_r(T) \ .
\label{mformdef}
\ee
One can show that there are no modular forms of weight 0 and 2 
while at weight 4 and 6 one has the Eisenstein functions
\bea
E_4 (q)& \equiv &  1 + 240 \sum_{n=1}^\infty \frac{n^3 q^n}{1-q^n} \nonumber\\
        &=& 1 + 240 q + 2160 q^2 \ldots\ , \\
E_6 (q) &\equiv&  1 - 504 \sum_{n=1}^\infty \frac{n^5 q^n}{1-q^n} \nonumber\\
        &=& 1 - 504 q -16632 q^2 \ldots\ , \nonumber
\label{Eisenstein}
\eea
where $q\equiv e^{-2\pi T}$.
Both function have no pole (including  $T=\infty$)
on the entire fundamental domain;
$E_4$ has exactly one simple zero at $T=\rho$ while 
$E_6$ has one simple zero at $T=1$. 
One can construct modular forms of arbitrary even weight from
products of these two Eisenstein functions.

A modular form which vanishes at $T=\infty$ is called a cusp form.
There is no cusp form of weight $r< 12$ and for $r=12$ there is the 
unique cusp form $\eta^{24}$ where 
\be
\eta(q) \equiv q^{1\over 24} \prod_{n=1}^\infty (1-q^n)
\label{etadef}
\ee
is the Dedekind $\eta$-function.
($\eta$ does not vanish at $\rho$ or 1.)

One can also construct a modular invariant function but it necessarily has 
a pole somewhere on the fundamental domain. The $j$--function defined by
\bea
j(q)& \equiv&  {E_4^3\over \eta^{24}} = {E_6^2\over \eta^{24}} + 1728\nonumber\\
   &  =& q^{-1} + 744 + 196884 q + \ldots
\label{jdef}
\eea
has a simple pole at $T=\infty$ and a triple zero at $T=\rho$.
This  function maps  the fundamental domain 
of $SL(2,{\bf Z})$
onto the complex plane.

In general the derivative of a modular form
is not a modular form since it does not
satisfy eq.~(\ref{modform}).
An exception is the derivative 
$\p_T^n F_{1-n}$ 
which transforms according to 
\be\label{modder}
\p_T^n F_{1-n}\to (icT+d)^{(n+1)}\p_T^n F_{1-n}
\ee
and thus is a modular form of weight $n+1$.

%
\section{Calabi-Yau manifolds}

In this appendix we briefly recall a few facts 
about Calabi--Yau manifolds which we frequently use in the main text.
(For a more extensive review see the lectures by R.~Plesser
or for example \cite{GSW,theisen1}.) 

A Calabi--Yau manifold $Y$  is a  Ricci-flat K\"ahler manifold
of vanishing first Chern-class. 
Its holonomy group is $SU(n)$ where $n$ is the complex
dimension of $Y$. 
The simplest 
Calabi--Yau manifolds  are tori of complex dimension 1.
For $n=2$ all Calabi--Yau manifolds are
topologically equivalent to the $K3$ surface (Kummers third surface),
while for $n=3$ one finds many topologically distinct
Calabi--Yau threefolds.
Such manifolds are of interest in string theory since
they break some of the supersymmetries when a ten-dimensional
string theory is compactified on $Y$.
(See table~4,5.)

The massless modes of a string vacuum are directly related
to the zero modes of the Laplace operator on  $Y$. 
These zero modes are 
the non-trivial differential $k$-forms 
on $Y$ and they are elements
of the cohomology groups $H^k(Y)$.
On a compact K\"ahler manifold one can decompose any $k$-form
into a $(p,q)$-form with $p$ holomorphic and $q$ antiholomorphic differentials.
Analogously, the associated cohomology groups
decompose according to
\be
H^k(Y)=\oplus_{p+q=k}\, H^{p,q}(Y).
\ee
The dimension of $H^{p,q}(Y)$ is called the
Hodge number $h_{p,q}$
($h_{p,q}={\rm dim}\, H^{p,q}$);
it is symmetric under the exchange of  
$p$ and $q$, i.e.~$h_{p,q}=h_{q,p}$, 
and Poincar{\'e} duality identifies
$h_{p,q}=h_{n-p,n-q}$. 
Finally, the Euler number is given  by 
\be\label{CYEuler}
\chi=\sum_{p,q}\, (-1)^{p+q}\, h_{p,q}\ .
\ee

For a $K3$ surface 
one finds for the independent
Hodge numbers $h_{0,0}=h_{2,0}=h_{2,2}=1$,
$h_{1,0}=h_{2,1}=0$ and $h_{1,1}=20$; hence 
eq.~(\ref{CYEuler}) implies $\chi(K3)=24$. 
For a Calabi--Yau threefold $Y$ the 
independent Hodge 
numbers obey 
$h_{0,0}=h_{3,0}=1$, $h_{1,0}=h_{2,0}=0$
while $h_{1,1}$ and $h_{1,2}$ are arbitrary. 
Thus eq.~(\ref{CYEuler}) implies 
$\chi(Y) = 2(h_{1,1} - h_{1,2})$. 
$h_{1,1}$ is the number of K\"ahler moduli, which are nontrivial deformations
of the  metric and the antisymmetric
tensor;
$h_{1,2}$ is the number of moduli, 
which are nontrivial deformations of the complex structure.
The moduli space is locally 
a direct product of the K\"ahler moduli space and 
the complex structure moduli space 
\be\label{cymodulisp}
{\cal M}={\cal M}_{h_{1,1}}\otimes{\cal M}_{h_{1,2}}\ .
\ee

It is believed that most Calabi--Yau threefolds (if not all)
have a mirror partner \cite{dixon,cls,GP}.
That is, for a given  Calabi--Yau threefold $Y$ with given 
$h_{1,1}(Y)$ and $h_{1,2}(Y)$ there exists a mirror manifold
 $\tilde Y$ with $h_{1,1}(\tilde Y) = h_{1,2}(Y)$
and $h_{1,2}(\tilde Y) = h_{1,1}(Y)$. (This implies
in particular $\chi(Y)= -\chi(\tilde Y)$.)

\end{appendix}
\bibliography{holcoupl}
\bibliographystyle{unsrt}
\end{document}